\documentclass[twocolumn]{aastex631}

\DeclareUnicodeCharacter{00A0}{ }

\usepackage{enumitem}
\usepackage{tabularx}
\usepackage{amsmath}
\usepackage{xstring}
\usepackage{xspace}
\usepackage{array}
\usepackage{multirow}
\usepackage{hhline}

\newcommand{\ergs}{\ensuremath{{\rm erg\ s}^{-1}}\xspace}
\newcommand{\sfrUnits}{\ensuremath{M_\odot/{\rm yr}}\xspace}

\newcommand{\Lx}{\ensuremath{L_{\rm X}}\xspace}
\newcommand{\logLx}{\ensuremath{\log\ L_{\rm X}}\xspace}
\newcommand{\logL}{\ensuremath{\log\ L}\xspace}
\newcommand{\LxOverSFR}{\ensuremath{L_{\rm X}/{\rm SFR}}\xspace}
\newcommand{\logoh}{\ensuremath{12\ +\ \log({\rm O/H})}\xspace}

\graphicspath{{./}{figures/}}

\shorttitle{The HMXB Luminosity Functions of Dwarf Galaxies}
\shortauthors{Geda et al.}

\begin{document}

\title{The High Mass X-ray Binary Luminosity Functions of Dwarf Galaxies}

\author[0000-0003-1509-9966]{Robel Geda}
\affiliation{Department of Astrophysical Sciences, Princeton University, Princeton, NJ 08540, USA}

\author[0000-0003-4700-663X]{Andy D Goulding}
\affiliation{Department of Astrophysical Sciences, Princeton University, Princeton, NJ 08540, USA}

\author[0000-0003-2192-3296]{Bret D. Lehmer}
\affiliation{Department of Physics, University of Arkansas, 226 Physics Building, 825 West Dickson Street, Fayetteville, AR 72701, USA}

\author[0000-0002-5612-3427]{Jenny E Greene}
\affiliation{Department of Astrophysical Sciences, Princeton University, Princeton, NJ 08540, USA}

\author[0000-0002-8477-7137]{Anish Kulkarni}
\affiliation{Department of Physics, Princeton University, Princeton, New Jersey 08540, USA}

\begin{abstract}

Drawing from the Chandra archive and using a carefully selected set of nearby dwarf galaxies, we present a calibrated high-mass X-ray binary (HMXB) luminosity function in the low-mass galaxy regime and search for an already hinted at dependence on metallicity. Our study introduces a new sample of local dwarf galaxies ($D < 12.5$ Mpc and $M_* < 5 \times 10^9\ M_\odot$), expanding the specific star-formation rates (sSFR) and gas-phase metallicities probed in previous investigations. Our analysis of the observed X-ray luminosity function indicates a shallower power-law slope for the dwarf galaxy HMXB population. In our study, we focus on dwarf galaxies that are more representative in terms of sSFR compared to prior work. In this regime, the HMXB luminosity function exhibits significant stochastic sampling at high luminosities. This likely accounts for the pronounced scatter observed in the galaxy-integrated HMXB population's \LxOverSFR versus metallicity for our galaxy sample. Our calibration is necessary to understand the AGN content of low mass galaxies identified in current and future X-ray survey fields and has implications for binary population synthesis models, as well as X-ray driven cosmic heating in the early universe.

\end{abstract}

\section{Introduction} 
\label{sec:intro}

High-mass X-ray binaries (HMXBs) are binary systems that consist of a compact object, a black hole or neutron star, that is accreting material from a high-mass companion star ($M_*>10 M_\odot$). Because of the relatively short lifetimes of the high-mass stars in HMXBs, the total X-ray luminosity of HMXBs serves as a powerful tool for probing star formation rates (SFRs) in galaxies \citep[e.g.,][]{Hornschemeier2000ApJ, Grimm2003MNRAS, Ranalli2003A&A, Persic2004A&A, Lehmer2010ApJ, Mineo2012MNRAS, Basu-Zych2013ApJ, Lehmer2016ApJ, Fornasini2018ApJ, Saxena2021MNRAS}. In addition, there is growing evidence suggesting that HMXBs in dwarf galaxies in the early universe may have contributed significantly to the ionizing radiation during the preheating of the intergalactic medium leading up to the reionization epoch \citep{Warszawski2009MNRAS, Madau2017ApJ, Eide2018MNRAS}. The study of HMXBs also provides an important constraint on binary star and compact object evolution, with important implications for gravitational wave sources \citep{Podsiadlowski2003MNRAS, Abbott2016ApJ, Liotine2023ApJ}. 

For galaxies that are actively forming stars, the HMXB X-ray luminosity function has been observed to predominantly correlate with the overall SFR of the host galaxy as expected \citep[][L19 hereafter]{Grimm2003MNRAS, Gilfanov2004MNRAS, Mineo2012MNRAS, L19}. However, recent observations and theoretical work suggest that the HMXB luminosity function may also depend on factors such as metallicity and star formation history (SFH). Various models have been proposed to describe the dependence of the luminosity function on metallicity \citep{Brorby2014MNRAS, Basu-Zych2016ApJ, Ponnada2020MNRAS}, primarily interpreting the \LxOverSFR versus metallicity relation. \cite{L21} (L21 hereafter) in particular observe that this metallicity dependence causes an excess of sources above $10^{38}$ \ergs for low-metallicity galaxies and introduce a framework for modeling the HMXB X-ray luminosity functions as a function of SFR and metallicity.

Thanks to their low metallicities \citep{Tremonti2004ApJ, Mannucci2010MNRAS}, dwarf galaxies serve as a valuable tool to examine the impact of low-metallicity environments on the HMXB X-ray luminosity function. In dwarf galaxies, another consideration becomes highly relevant. HMXBs are the dominant contributors to the overall X-ray luminosity of any star-forming galaxy without an active galactic nucleus (AGN). In dwarf galaxies, however, the contribution of HMXBs may dominate over that of an active nucleus as well, since dwarf galaxies may harbor very low-mass central black holes \citep{Mezcua2017IJMPD, Greene2020ARA&A}. The X-ray emission from HMXBs becomes a dominant source of confusion in the context of intermediate-mass black hole (IMBH) detection in distant dwarf galaxies, where we cannot resolve individual HMXBs with current telescopes \citep{Schramm2013ApJ, Pardo2016ApJ, Mezcua2018MNRAS, Halevi2019ApJ}. Without a better understanding of the full range of \Lx per unit star-formation rate from dwarfs, it is not possible to robustly identify AGN candidates using X-rays; while there is a hope to detect and characterize seed black holes in the early universe using next-generation X-ray missions \citep{Natarajan2017ApJ, Barrow2018MNRAS, Ricarte2018MNRAS, Haiman2019BAAS}. A better handle on the role of metallicity in setting the HMXB luminosity function is urgently needed for this purpose.

In this work, we seek to understand this possible metallicity dependence more securely by increasing the sample of dwarf galaxies with measured HMXB luminosity functions. Prior studies suggest that local dwarf galaxies exhibit a metallicity-driven surplus of high-\Lx sources, consequently affecting the shape of their X-ray luminosity functions \citep{Mapelli2010MNRAS, prestwich_2013, Brorby2014MNRAS, Douna2015A&A, Kovlakas2020MNRAS}. In particular, there appears to be an excess of so-called ``ultra-luminous X-ray'' (ULX) sources in low-mass and low-metallicity galaxies. ULXs are X-ray point sources with luminosities that exceed the Eddington limit for stellar-mass black holes ($10^{39}$ \ergs). Fully characterizing the HMXB X-ray luminosity function for a larger and more representative sample of dwarf galaxies will quantify whether low metallicity leads to this apparent ULX excess. Further, as shown by \cite{Fornasini2020MNRAS}, the metallicity dependence of \LxOverSFR is not expected to evolve with redshift, allowing us to consider local dwarfs as useful analogs for distant dwarf galaxies.

Past studies have often focused on a limited sample of dwarf galaxies, predominantly favoring those with high specific star-formation rates and the lowest known metallicities \citep{prestwich_2013}. L21 in particular relies on these galaxies to constrain their low-metallicity luminosity functions, but leave a notable gap between $\logoh \approx 7.6-8.0$ range as a consequence. We fill this gap by broadening our sample to include a wider metallicity range of $\logoh=7.74-8.77$, thus encompassing typical dwarfs with $M_* \sim 10^8-10^9~M_\odot$ and $\log$ sSFR $\sim -10.7 - -8.5$. In addition to being targeted by the Chandra X-ray Observatory, these galaxies also benefit from supporting data from GALEX and HST, enabling us to define uniform apertures and obtain reliable measurements of their SFRs. 

The structure of this paper is as follows: In Section \ref{sec:sample}, we discuss our selected sample of galaxies. Section \ref{sec:data} outlines the data preparation for our analysis. The luminosity function models and the results of our fitting are detailed in Sections \ref{sec:modeling} and \ref{sec:luminosity_fits}, respectively. Finally, Section \ref{sec:lx-sfr-z} discusses the \Lx-SFR-metallicity relation observed in our sample.

\section{Galaxy Sample} 
\label{sec:sample}

\begin{figure}[t]
\includegraphics[width=8cm]{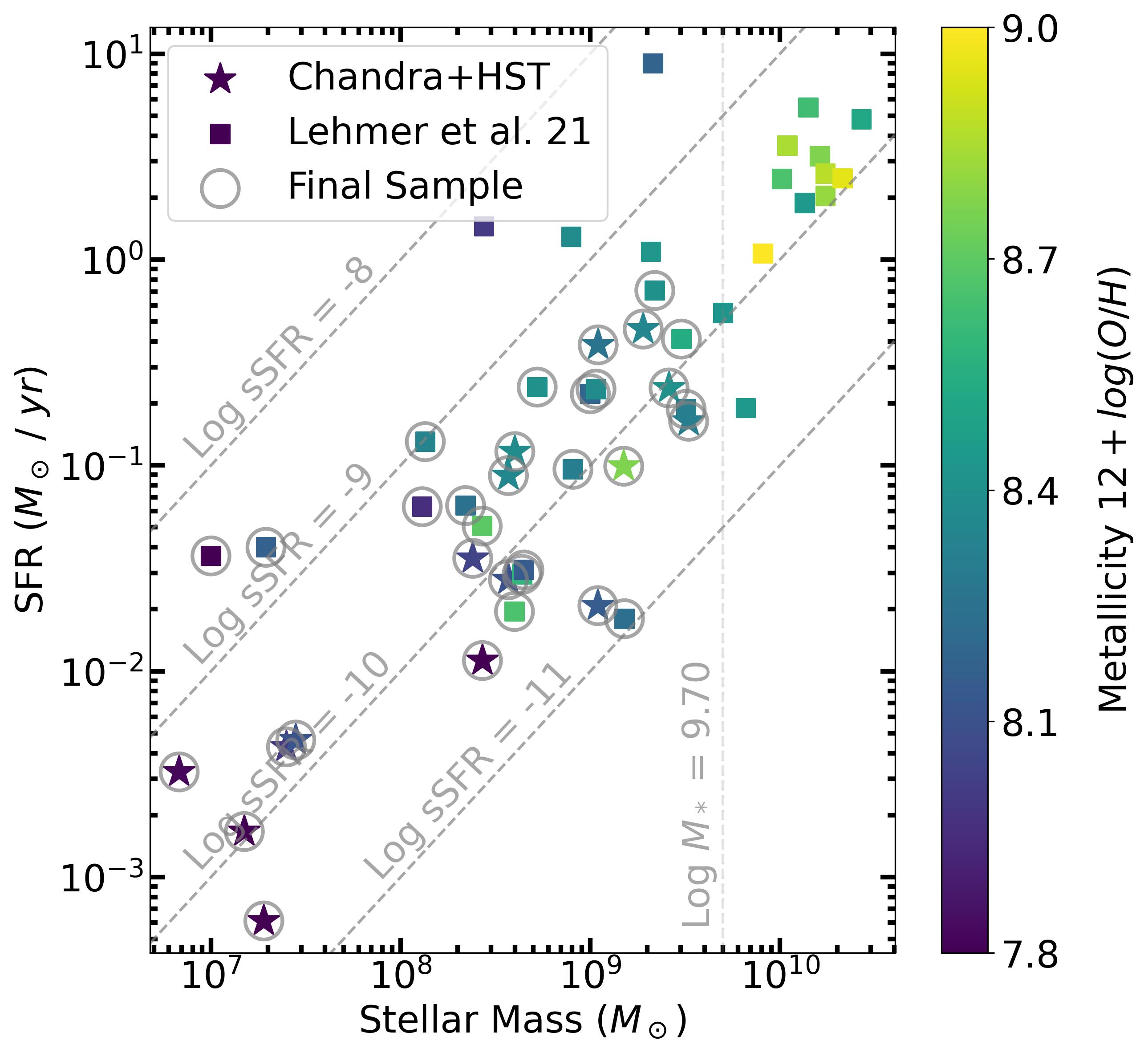}
\centering
\caption{\label{fig:sfr_vs_mass} 
Stellar mass versus SFR for the galaxies in the L21 galaxy sample (``Lehmer et al. 21") and galaxies introduced in this study (``Chandra+HST"). Chandra+HST are reprinted by stars while L21 galaxies are marked by square points. The points are colored according to gas-phase metallicity (\logoh). Our dwarf galaxy sample supplements the L21 sample in $M_{*}$, SFR, and metallicity while maintaining a comparable sSFR. The gray circles denote galaxies that satisfy the mass ($M_{*} < 5 \times 10^9$~Msun) and distance ($D<12$~Mpc) requirements (in both samples), mainly set by limits in STARBIRDS and LEGUS that our sample is drawn from. The diagonal gray lines show log sSFRs and the vertical line denotes our mass cutoff limit. 
}
\end{figure}

Here we present the ``Chandra+HST'' dwarf sample. We focus on local dwarf galaxies with stellar masses $M_* < 5 \times 10^9\ M_\odot$ from two Local Volume surveys ($D<12.5$~Mpc). The Chandra data are used to identify high-mass X-ray binaries, while the HST data ensure high-fidelity mass and distance measurements for the galaxies. Specifically, we start with a primary sample consisting of two large complementary sets of nearby galaxies, the Legacy ExtraGalactic UV Survey  \citep[LEGUS;][]{LEGUS_1, LEGUS_3, LEGUS_2} and the STARBurst IRregular Dwarf Survey \citep[STARBIRDS;][]{STARBIRDS_2015}. Both LEGUS and STARBIRDS have sensitive multi-band coverage from the Hubble Space Telescope's (HST) WFC3, Galaxy Evolution Explorer Telescope \citep[][GALEX]{GALEX}, and the Spitzer Space Telescope's IRAC offering full UV to IR coverage of the galaxy spectral energy distributions.  Given the high quality of available data for these sources, they each have extremely accurate (1) direct distances derived from tip of the red giant branch (TRGB), Surface Brightness Fluctuations, Cepheids, and/or SNeII measurements; (2) $M_*$ measurements using multi-band photometry; (3) SFRs from HST+GALEX photometry; and (4) gas-phase metallicity measurements using oxygen abundances obtained from spectroscopic follow-up. We use X-ray data from NASA's Chandra X-ray Observatory, which offers the angular resolution and sensitivity needed to quantify the X-ray binary (XRB) populations in nearby galaxies. Among the 52 galaxies in the STARBIRDS and LEGUS datasets with stellar masses $M_* < 5 \times 10^9 M_\odot$, $30$ ($58\%$) of these galaxies have publicly available data from Chandra. These observations were taken using the Advanced CCD Imaging Spectrometer (ACIS) from either I (Imaging) or S (Spectroscopy) cameras. Specifically, we label the sample of galaxies that have Chandra ACIS-I/S data as the Chandra+HST sample.

The stellar masses for the Chandra+HST sample were measured by \cite{calzetti2015_mass} and \cite{mcquinn2010_table} for the LEGUS and STARBIRDS sub-samples, respectively. \cite{calzetti2015_mass} obtain stellar masses from extinction-corrected B-band luminosities and color information, using the method described in \cite{bothwell2009} and based on the mass-to-light ratio models from \cite{bell2001}. \cite{mcquinn2010_table} calculate the total amount of stellar mass from published absolute B-band luminosities of the galaxies, adjusted for extinction from Galactic dust maps published by \cite{schlegel1998}. We use metallicities compiled by \cite{calzetti2015_mass} based on direct temperature measurements in the literature. For the Chandra+HST sample, we use distances reported by \cite{lee2009_tables}. Galaxies in the Chandra+HST sample span a mass range of $\text{log}\ (M_*/M_\odot) = 6.8-9.5$ and a distance range of $0.5-12.5$ Mpc.

Although $30$ galaxies in LEGUS+Starbirds have Chandra data, not all of them have sufficient data for our purposes. We remove two galaxies that were only observed with a sub-array, which is unable to cover the full extent of our nearby galaxies. We then remove an additional six galaxies because they do not reach our required depth of $\logL\ (\ergs) > 37.5$ (see Section \ref{subsec:rf} for details). 

The rejected galaxies and the reasons that they were rejected are indicated by flags in the Flag column of Table \ref{table:overview}. The total number of Chandra+HST dwarf galaxies in the sample after all cuts is 22.

\begin{figure*}[t]
\includegraphics[width=\textwidth]{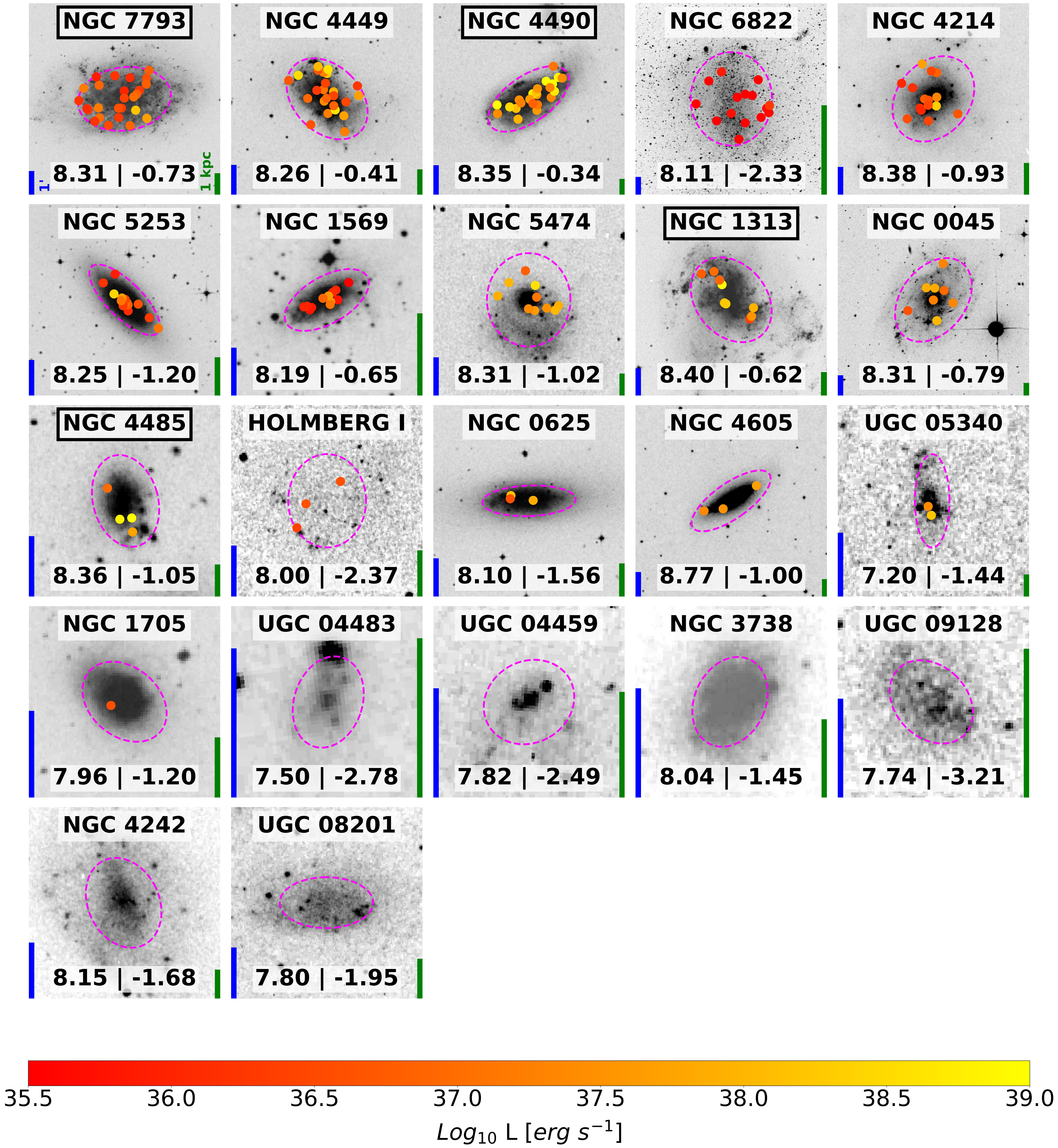}
\centering
\caption{\label{fig:galaxies}
This figure shows the dwarf galaxies in the Chandra+HST final sample. In each panel, we include a DSS image of the galaxy, which is displayed in gray-scale. Galaxy names are displayed at the top of each panel and the corresponding [\logoh, log(SFR)] values are displayed at the bottom. If the galaxy was targeted by Chandra for ULXs, the galaxy is noted with a black box around its name. The apertures used in this study are plotted as ellipses (magenta). X-ray sources are overplotted as circles and colored according to their X-ray luminosities. For reference, vertical bars of size  1' (blue) and 1 kpc at the galaxy's distance (green) are provided in the lower left and lower right corners of each panel, respectively.}
\end{figure*}

\begin{flushleft}
\begin{table*}[t]
\footnotesize
\centering
\begin{tabular}{@{}l *{11}{c} l@{}}
\hline \hline
\multicolumn{1}{c}{Name} & \multicolumn{1}{c}{$\alpha_{J2000}$} & \multicolumn{1}{c}{$\delta_{J2000}$} & \multicolumn{1}{c}{E(B-V)} & \multicolumn{1}{c}{D} & \multicolumn{1}{c}{a} & \multicolumn{1}{c}{b} & \multicolumn{1}{c}{PA} & \multicolumn{1}{c}{M} & \multicolumn{1}{c}{12+log[O/H]} & \multicolumn{1}{c}{FUV Mag} & \multicolumn{1}{c}{Flag} & \multicolumn{1}{c}{Source}\\ 
\multicolumn{1}{c}{(1)} & \multicolumn{1}{c}{(2)} & \multicolumn{1}{c}{(3)} & \multicolumn{1}{c}{(4)} & \multicolumn{1}{c}{(5)} & \multicolumn{1}{c}{(6)} & \multicolumn{1}{c}{(7)} & \multicolumn{1}{c}{(8)} & \multicolumn{1}{c}{(9)} & \multicolumn{1}{c}{(10)} & \multicolumn{1}{c}{(11)} & \multicolumn{1}{c}{(12)} & \multicolumn{1}{c}{(13)}\\ 
\hline
  & H M S & D M S &   & Mpc & arcmin & arcmin & Deg & $M_{\odot}$ &   & Mag &   &  \\ 
\hline
NGC 0045 & 00 14 04.0 & -23 10 55.0 & $0.02$ & $7.1$ & $3.15$ & $2.18$ & $-38$ & $9.52$ & $8.31$ & $13.18 \pm 0.08$ & - & LG\\ 
NGC 0625 & 01 35 04.2 & -41 26 15.0 & $0.02$ & $4.1$ & $1.43$ & $0.47$ & $-88$ & $8.57$ & $8.10$ & $13.90 \pm 0.12$ & - & SB\\ 
IC 1727 & 01 47 30.6 & +27 19 52.0 & $0.08$ & $7.2$ & $1.27$ & $0.57$ & $-30$ & $8.74$ & $8.73$ & - & RF & LG\\ 
NGC 1313 & 03 18 15.8 & -66 29 53.0 & $0.11$ & $4.2$ & $2.15$ & $1.63$ & $+40$ & $9.41$ & $8.40$ & $11.61 \pm 0.06$ & - & LG\\ 
NGC 1569 & 04 30 49.0 & +64 50 53.0 & $0.69$ & $1.9$ & $1.10$ & $0.55$ & $-60$ & $9.00$ & $8.19$ & $9.98 \pm 0.20$ & - & SB+L21\\ 
NGC 1705 & 04 54 13.7 & -53 21 41.0 & $0.01$ & $5.1$ & $0.57$ & $0.42$ & $+50$ & $8.11$ & $7.96$ & $13.50 \pm 0.09$ & - & LG+L21\\ 
NGC 2500 & 08 01 53.3 & +50 44 15.0 & $0.04$ & $7.6$ & $0.88$ & $0.78$ & $0$ & $9.28$ & $8.84$ & - & RF & LG\\ 
UGC 04459 & 08 34 07.2 & +66 10 54.0 & $0.04$ & $3.6$ & $0.45$ & $0.39$ & $-60$ & $6.83$ & $7.82$ & $15.94 \pm 0.29$ & - & LG\\ 
UGC 04483 & 08 37 03.0 & +69 46 31.0 & $0.03$ & $3.2$ & $0.33$ & $0.24$ & $-18$ & $7.18$ & $7.50$ & $16.44 \pm 0.35$ & - & SB\\ 
Holmberg I & 09 40 32.3 & +71 10 56.0 & $0.05$ & $3.8$ & $1.02$ & $0.85$ & $0$ & $7.40$ & $8.00$ & $15.80 \pm 0.30$ & - & LG\\ 
UGC 05340 & 09 56 45.7 & +28 49 35.0 & $0.02$ & $12.1$ & $0.80$ & $0.30$ & $0$ & $7.00$ & $7.20$ & $15.97 \pm 0.28$ & - & LG+L21\\ 
NGC 3274 & 10 32 17.1 & +27 40 07.0 & $0.02$ & $6.5$ & $1.02$ & $0.49$ & $-80$ & $8.04$ & $8.33$ & - & RF & LG\\ 
UGC 06456 & 11 28 00.0 & +78 59 39.0 & $0.04$ & $4.3$ & $0.42$ & $0.24$ & $-10$ & $7.83$ & $7.64$ & - & SA & SB\\ 
NGC 3738 & 11 35 48.8 & +54 31 26.0 & $0.01$ & $4.9$ & $0.45$ & $0.34$ & $-25$ & $8.38$ & $8.04$ & $14.04 \pm 0.12$ & - & LG\\ 
NGC 4214 & 12 15 38.9 & +36 19 40.0 & $0.02$ & $2.9$ & $2.10$ & $1.63$ & $-40$ & $8.60$ & $8.38$ & $11.62 \pm 0.04$ & - & SB\\ 
NGC 4242 & 12 17 30.1 & +45 37 08.0 & $0.01$ & $7.4$ & $0.92$ & $0.70$ & $+25$ & $9.04$ & $8.15$ & $15.51 \pm 0.23$ & - & LG\\ 
NGC 4395 & 12 25 48.9 & +33 32 48.0 & $0.02$ & $4.6$ & $2.45$ & $2.04$ & $-33$ & $8.77$ & $8.26$ & - & SA & LG\\ 
UGCA 281 & 12 26 16.0 & +48 29 37.0 & $0.01$ & $5.7$ & $0.28$ & $0.21$ & $-85$ & $7.28$ & $7.82$ & $15.53 \pm 0.22$ & RF & LG\\ 
NGC 4449 & 12 28 11.2 & +44 05 36.0 & $0.02$ & $4.2$ & $1.93$ & $1.37$ & $+45$ & $9.04$ & $8.26$ & $11.12 \pm 0.03$ & - & LG\\ 
NGC 4485 & 12 30 31.1 & +41 42 01.0 & $0.02$ & $7.1$ & $0.85$ & $0.59$ & $+15$ & $8.57$ & $8.36$ & $13.84 \pm 0.11$ & - & LG\\ 
NGC 4490 & 12 30 36.1 & +41 38 34.0 & $0.02$ & $8.0$ & $1.95$ & $0.96$ & $-55$ & $9.28$ & $8.35$ & $12.32 \pm 0.06$ & - & LG\\ 
NGC 4605 & 12 40 00.3 & +61 36 29.0 & $0.01$ & $5.5$ & $2.45$ & $0.93$ & $-55$ & $9.18$ & $8.77$ & $13.16 \pm 0.08$ & - & LG\\ 
UGC 08201 & 13 06 24.8 & +67 42 25.0 & $0.02$ & $4.6$ & $1.02$ & $0.56$ & $+90$ & $8.43$ & $7.80$ & $15.13 \pm 0.20$ & - & SB\\ 
NGC 5253 & 13 39 55.9 & -31 38 24.0 & $0.06$ & $3.1$ & $1.55$ & $0.59$ & $+45$ & $8.34$ & $8.25$ & $12.44 \pm 0.07$ & - & LG+L21\\ 
NGC 5474 & 14 05 01.5 & +53 39 45.0 & $0.01$ & $6.8$ & $1.43$ & $1.27$ & $0$ & $8.91$ & $8.31$ & $13.67 \pm 0.10$ & - & LG+L21\\ 
UGC 09128 & 14 15 56.5 & +23 03 19.0 & $0.02$ & $2.2$ & $0.50$ & $0.38$ & $+45$ & $7.28$ & $7.74$ & $16.74 \pm 0.38$ & - & SB\\ 
NGC 5949 & 15 28 00.7 & +64 45 47.0 & $0.02$ & $8.5$ & $0.68$ & $0.31$ & $-33$ & $9.26$ & $8.37$ & - & RF & LG\\ 
NGC 6503 & 17 49 27.1 & +70 08 40.0 & $0.03$ & $5.3$ & $2.20$ & $0.74$ & $-57$ & $9.28$ & $8.51$ & $13.64 \pm 0.11$ & RF & LG\\ 
NGC 6822 & 19 44 56.6 & -14 47 21.0 & $0.23$ & $0.5$ & $3.82$ & $3.33$ & $0$ & $7.45$ & $8.11$ & $11.29 \pm 0.09$ & - & SB\\ 
NGC 7793 & 23 57 49.7 & -32 35 30.0 & $0.02$ & $3.9$ & $2.58$ & $1.74$ & $-82$ & $9.51$ & $8.31$ & $11.74 \pm 0.05$ & - & LG+L21\\ 
\\[-0.3cm] 
\hline 
\\[-0.3cm] 
NGC 0024 & 00 09 56.7 & -24 57 44.0 & $0.02$ & $7.3$ & $1.07$ & $0.24$ & $+46$ & $8.64$ & $8.59$ & $15.10 \pm 0.20$ & - & L21\\ 
MESSIER 074 & 01 36 41.7 & +15 46 59.0 & $0.07$ & $7.3$ & $3.25$ & $2.94$ & $+25$ & $9.48$ & $8.54$ & $12.24 \pm 0.07$ & - & L21\\ 
NGC 0925 & 02 27 16.9 & +33 34 45.0 & $0.08$ & $9.1$ & $1.90$ & $1.07$ & $-78$ & $9.03$ & $8.38$ & $13.33 \pm 0.12$ & - & L21\\ 
ESO 495- G 021 & 08 36 15.2 & -26 24 33.7 & - & $9.0$ & $0.89$ & $0.80$ & $+140$ & $8.72$ & $8.40$ & - & - & L21\\ 
NGC 2915 & 09 26 11.5 & -76 37 36.0 & $0.27$ & $3.8$ & $0.57$ & $0.30$ & $-51$ & $8.65$ & $8.15$ & $13.62 \pm 0.25$ & - & L21\\ 
NGC 2976 & 09 47 15.3 & +67 55 00.0 & $0.07$ & $3.6$ & $1.10$ & $0.50$ & $-37$ & $8.60$ & $8.66$ & $13.99 \pm 0.15$ & - & L21\\ 
NGC 3125 & 10 06 33.6 & -29 56 09.0 & $0.08$ & $12.0$ & $0.53$ & $0.33$ & $-66$ & $8.13$ & $8.34$ & $14.57 \pm 0.19$ & - & L21\\ 
IC 2574 & 10 28 21.2 & +68 24 43.0 & $0.04$ & $4.0$ & $2.45$ & $1.00$ & $+50$ & $9.18$ & $8.23$ & $14.34 \pm 0.16$ & - & L21\\ 
NGC 4559 & 12 35 57.7 & +27 57 35.1 & $0.02$ & $10.3$ & $2.04$ & $0.96$ & $+148$ & $9.34$ & $8.40$ & $12.40 \pm 0.06$ & - & L21\\ 
NGC 4625 & 12 41 52.6 & +41 16 26.0 & $0.02$ & $9.2$ & $1.00$ & $0.86$ & $-30$ & $8.43$ & $8.70$ & $15.02 \pm 0.19$ & - & L21\\ 
NGC 5408 & 14 03 20.9 & -41 22 39.8 & - & $4.8$ & $0.75$ & $0.75$ & $+95$ & $7.29$ & $8.17$ & - & - & L21\\ 
\hline 
\end{tabular}

\caption{The Chandra+HST sample before selection and selected L21 dwarfs (separated by horizontal line). (1) Galaxy name. (2)-(3) Central position RA and Dec. (4) E(B-V) Milky-Way extinction. (5) Distance from \cite{lee2009_tables} for Chandra Dwarfs and \cite{L21} for L21 galaxies. (6)-(7) Semi-major and semi-minor radii of galaxy aperture used in this study. (8) Position angle of aperture. (9) Log stellar mass. (10) Gas-phase metallicity. (11) GALEX FUV Mag, total flux within galaxy aperture. (12) Cause of rejection for galaxies that were not included in the analysis where M is mass cutoff, SA is sub-array mode, and RF is recovery function cutoff. (13) Source catalog where LG is LEGUS, SB is STARBIRDS, and L21.}\label{table:overview}
\end{table*}
\end{flushleft}

\begin{flushleft}
\begin{table*}[t]
\footnotesize
\centering
\begin{tabular}{l*{11}{c}}
\hline \hline
\multicolumn{1}{c}{Name} & \multicolumn{1}{c}{ObsID} & \multicolumn{1}{c}{Cycle} & \multicolumn{1}{c}{t} & \multicolumn{1}{c}{ACIS} & \multicolumn{1}{c}{Log SFR} & \multicolumn{1}{c}{N} & \multicolumn{1}{c}{$L_x$} & \multicolumn{1}{c}{$L_{CXB}$} & \multicolumn{1}{c}{$L_{LMXB}$} & \multicolumn{1}{c}{$L_{Peak}$} & \multicolumn{1}{c}{ULX}\\ 
\multicolumn{1}{c}{(1)} & \multicolumn{1}{c}{(2)} & \multicolumn{1}{c}{(3)} & \multicolumn{1}{c}{(4)} & \multicolumn{1}{c}{(5)} & \multicolumn{1}{c}{(6)} & \multicolumn{1}{c}{(7)} & \multicolumn{1}{c}{(8)} & \multicolumn{1}{c}{(9)} & \multicolumn{1}{c}{(10)} & \multicolumn{1}{c}{(11)} & \multicolumn{1}{c}{(12)}\\ 
\hline
  &   &   & Ks &   & log $yr^{-1}$ &   & log erg $s^{-1}$ & log erg $s^{-1}$ & log erg $s^{-1}$ & log erg $s^{-1}$ &  \\ 
\hline
NGC 0045 & $4690$ & $5$ & $34.40$ & $S$ & $-0.79 \pm 0.032$ & $8$ & $38.51 \pm 0.028$ & $38.7$ & $38.6$ & $38.0$ & $N$\\ 
NGC 0625 & $4746$ & $5$ & $60.30$ & $S$ & $-1.56 \pm 0.047$ & $3$ & $38.35 \pm 0.014$ & $37.3$ & $37.7$ & $38.2$ & $N$\\ 
NGC 1313 & $2950$ & $3$ & $19.90$ & $S$ & $-0.62 \pm 0.024$ & $9$ & $39.57 \pm 0.006$ & $37.9$ & $38.5$ & $39.5$ & $Y$\\ 
NGC 1569 & $782$ & $1$ & $95.00$ & $S$ & $-0.65 \pm 0.080$ & $12$ & $37.99 \pm 0.009$ & $36.6$ & $38.1$ & $37.6$ & $N$\\ 
NGC 1705 & $3930$ & $4$ & $56.00$ & $S$ & $-1.20 \pm 0.036$ & $1$ & $36.60 \pm 0.188$ & $37.0$ & $37.2$ & $36.6$ & $N$\\ 
UGC 04459 & $9538$ & $9$ & $25.90$ & $S$ & $-2.49 \pm 0.116$ & $0$ & $0.00 \pm 0.000$ & $36.5$ & $36.0$ & $0.0$ & $N$\\ 
UGC 04483 & $10559$ & $10$ & $3.09$ & $S$ & $-2.78 \pm 0.140$ & $0$ & $0.00 \pm 0.000$ & $35.9$ & $36.3$ & $0.0$ & $N$\\ 
Holmberg I & $9539$ & $9$ & $25.90$ & $S$ & $-2.37 \pm 0.120$ & $3$ & $37.03 \pm 0.088$ & $37.3$ & $36.5$ & $36.6$ & $N$\\ 
UGC 05340 & $11271$ & $11$ & $118.00$ & $S$ & $-1.44 \pm 0.112$ & $2$ & $38.26 \pm 0.043$ & $37.8$ & $36.1$ & $38.2$ & $N$\\ 
NGC 3738 & $19357$ & $18$ & $9.34$ & $S$ & $-1.45 \pm 0.048$ & $0$ & $0.00 \pm 0.000$ & $36.7$ & $37.5$ & $0.0$ & $N$\\ 
NGC 4214 & $2030$ & $2$ & $26.40$ & $S$ & $-0.93 \pm 0.016$ & $15$ & $38.49 \pm 0.012$ & $37.6$ & $37.7$ & $38.3$ & $N$\\ 
NGC 4242 & $19351$ & $18$ & $9.94$ & $S$ & $-1.68 \pm 0.092$ & $0$ & $0.00 \pm 0.000$ & $37.7$ & $38.1$ & $0.0$ & $N$\\ 
NGC 4449 & $10875$ & $10$ & $59.40$ & $S$ & $-0.41 \pm 0.014$ & $24$ & $39.24 \pm 0.005$ & $37.9$ & $38.2$ & $38.6$ & $N$\\ 
NGC 4485 & $4726$ & $5$ & $39.60$ & $S$ & $-1.05 \pm 0.044$ & $4$ & $39.44 \pm 0.009$ & $37.6$ & $37.7$ & $39.3$ & $Y$\\ 
NGC 4490 & $4725$ & $5$ & $38.50$ & $S$ & $-0.34 \pm 0.024$ & $23$ & $39.98 \pm 0.005$ & $38.3$ & $38.4$ & $39.4$ & $Y$\\ 
NGC 4605 & $19344$ & $18$ & $9.67$ & $S$ & $-1.00 \pm 0.032$ & $3$ & $38.02 \pm 0.081$ & $38.0$ & $38.3$ & $37.7$ & $N$\\ 
UGC 08201 & $9537$ & $9$ & $13.50$ & $S$ & $-1.95 \pm 0.080$ & $0$ & $0.00 \pm 0.000$ & $37.2$ & $37.5$ & $0.0$ & $N$\\ 
NGC 5253 & $2032$ & $2$ & $190.00$ & $S$ & $-1.20 \pm 0.028$ & $15$ & $38.55 \pm 0.007$ & $37.2$ & $37.5$ & $38.4$ & $N$\\ 
NGC 5474 & $9546$ & $9$ & $31.00$ & $S$ & $-1.02 \pm 0.040$ & $10$ & $38.94 \pm 0.022$ & $38.1$ & $38.0$ & $38.6$ & $N$\\ 
UGC 09128 & $16121$ & $15$ & $4.91$ & $I$ & $-3.21 \pm 0.154$ & $0$ & $0.00 \pm 0.000$ & $35.7$ & $36.3$ & $0.0$ & $N$\\ 
NGC 6822 & $2925$ & $3$ & $28.10$ & $I$ & $-2.33 \pm 0.036$ & $15$ & $36.72 \pm 0.018$ & $36.6$ & $36.6$ & $36.3$ & $N$\\ 
NGC 7793 & $3954$ & $4$ & $190.00$ & $S$ & $-0.73 \pm 0.020$ & $26$ & $38.51 \pm 0.009$ & $38.1$ & $38.6$ & $38.2$ & $Y$\\ 
\\[-0.3cm] 
\hline 
\\[-0.3cm] 
NGC 0024 & - & - & $43.00$ & - & $-1.53 \pm 0.078$ & $6$ & $38.03 \pm 0.054$ & $37.4$ & $37.8$ & $37.5$ & $N$\\ 
MESSIER 074 & - & - & $268.00$ & - & $-0.39 \pm 0.028$ & $79$ & $39.43 \pm 0.005$ & $39.0$ & $38.6$ & $39.1$ & $N$\\ 
NGC 0925 & - & - & $12.00$ & - & $-0.63 \pm 0.046$ & $7$ & $39.86 \pm 0.017$ & $38.4$ & $38.1$ & $39.6$ & $N$\\ 
ESO 495- G 021 & - & - & $216.00$ & - & $-0.62$ & $9$ & $38.73 \pm 0.014$ & $38.0$ & $37.8$ & $38.4$ & $N$\\ 
NGC 2915 & - & - & $15.00$ & - & $-1.51 \pm 0.100$ & $0$ & $0.00 \pm 0.000$ & $36.6$ & $37.8$ & $0.0$ & $N$\\ 
NGC 2976 & - & - & $9.00$ & - & $-1.71 \pm 0.058$ & $2$ & $38.80 \pm 0.021$ & $37.0$ & $37.7$ & $38.8$ & $N$\\ 
NGC 3125 & - & - & $64.00$ & - & $-0.88 \pm 0.077$ & $4$ & $39.61 \pm 0.012$ & $37.6$ & $37.2$ & $39.5$ & $N$\\ 
IC 2574 & - & - & $11.00$ & - & $-1.74 \pm 0.063$ & $2$ & $37.76 \pm 0.179$ & $37.7$ & $38.3$ & $37.7$ & $N$\\ 
NGC 4559 & - & - & $22.00$ & - & $-0.15 \pm 0.024$ & $5$ & $40.09 \pm 0.009$ & $38.5$ & $38.4$ & $40.0$ & $N$\\ 
NGC 4625 & - & - & $56.00$ & - & $-1.30 \pm 0.076$ & $7$ & $38.15 \pm 0.041$ & $38.1$ & $37.5$ & $37.9$ & $N$\\ 
NGC 5408 & - & - & $70.00$ & - & $-1.40$ & $7$ & $39.61 \pm 0.004$ & $37.4$ & $36.4$ & $39.6$ & $N$\\ 
\hline 
\end{tabular}
\caption{Our final sample (L21 dwarfs separated by horizontal line). (1) Galaxy Name. (2)  Chandra observation ID. (3) Chandra observation cycle. (4) Chandra exposure time. (5) ACIS configuration I for Wide Field Imaging and S for S3 chips. (6) Log SFR derived from FUV measurement. If FUV information is not available for L21 galaxies the value from L21 is used. (7) Number of Lx sources within aperture. (8) Total point source \Lx. (9) Total recovery function corrected \Lx from CXB \citep{kim_cxb_2007}. (10) Total, recovery function corrected, \Lx expected from LMXB (L19). (11) \Lx of the most luminous point source within aperture. (12) Yes if the galaxy was targeted for ULXs.}\label{table:photo}
\end{table*}
\end{flushleft}

\subsection{Additional Dwarfs from Lehmer et al. 2021 (L21)}
\label{subsec:l21_sample}

In addition to our primary sample, after applying our mass ($M_*<5 \times 10^9$~$M_{\odot}$) and distance ($D<12.5$~Mpc) cuts, we obtain 11 local dwarfs from the L21 main sample that are not in the Chandra+HST sample. We identified six additional dwarf galaxies (UGC 05340, NGC 1705, NGC 1569, NGC 5253, NGC 5474, and NGC 7793) that are in both L21 and Chandra+HST. We list the overlapping galaxies under the Chandra+HST category to avoid duplication. For these galaxies we use X-ray photometry and distances from L21 to preserve the X-ray luminosities reported in that work, and any other galaxy properties from measurements made for the Chandra+HST sample (see Section \ref{sec:data}). 

The galaxies in the original L21 main sample span a mass range of $\text{log}\ (M_*/M_\odot) = 7.3-10.4$ and a distance range of $1.9-29.4$~Mpc. The metallicities of galaxies in the L21 sample were derived using oxygen abundance measurements either from strong-line calibrations or direct electron–temperature-based theoretical calibration \citep{L21}.  It is worth noting that for objects in common between the two samples, the metallicities are in good agreement despite being derived from different sources.

\subsection{ULX Galaxies}

Our goal is an unbiased dwarf sample to study X-ray point source distributions. If galaxies were preferentially targeted by Chandra because of known bright X-ray point sources (i.e., ULXs), this could bias the number of detected sources at the bright end of the luminosity function. Many galaxies in the sample were targeted because of the known presence of a ULX. Specifically, we find that four galaxies -- NGC 7793, NGC 4490, NGC 4485, and NGC 1313 -- were targeted for ULXs. Therefore, we fit the X-ray luminosity functions with and without the ULX-targeted sample to assess the potential bias introduced by their addition.  

\subsection{Summary of Final Sample}
\label{subsec:final_sample}

We have identified a total of 33 individual galaxies that meet our criteria, 22 from Chandra+HST and 11 from L21. We will refer to this set of galaxies as our final sample. However, we further perform our analysis with and without the four galaxies targeted because they harbor ULXs. The sample containing ULX galaxies will be referred to as the ``final+ULX'' sample. The galaxies in the final sample span a stellar mass range of $\text{log}\ (M_*/M_\odot) = 6.8-9.52$, gas-phase metallicity range of $\logoh = 7.74-8.77$, and a distance range of $0.5-12.1$~Mpc. 

\section{Data Analysis} 
\label{sec:data}

In this section, we discuss galaxy apertures (Section \ref{subsec:aperture}), star formation rate (SFR) measurements (Section \ref{subsec:sfr}), X-ray photometry (Section \ref{subsec:x-rayphoto}), and recovery (completeness) functions (Section \ref{subsec:rf}) used in this study. We compute galaxy sizes, recovery functions, and perform X-ray photometry for all galaxies in the Chandra+HST sample (i.e before final sample selection) while utilizing published X-ray catalogs from L21 for galaxies covered in that study. We compute galaxy sizes and associated FUV SFRs for galaxies in the final sample. Table \ref{table:photo} presents X-ray and SFR measurements for galaxies that met all of our selection criteria (i.e final sample + ULX).

\subsection{Galaxy Apertures Using GALEX}
\label{subsec:aperture}

We use galaxy projected footprints as apertures when making photometric measurements and as boundaries when filtering for X-ray point sources. L21 uses apertures defined by 2MASS \citep{2mass} $20\ mag\ arcsec^{-2}$ isophotal ellipses in the $K_s$-band. However, several dwarf galaxies in this study are fainter in the $K_s$-band than the L21 sample because they are relatively blue, causing the 2MASS $K_s$-band apertures to severely underestimate the sizes of those galaxies, and therefore underestimate the number of HMXBs within them. Taking this into account, we use elliptical apertures from GALEX as a basis to define the projected apertures of galaxies in this study. 

In their study, \cite{lee2009_tables} define the semi-major and semi-minor axes of the outermost elliptical annulus for each galaxy from the GALEX photometry. This boundary is determined based on two criteria: it is either the point where the annular flux error exceeds $0.8$ magnitudes or the point where the intensity drops below the level of the sky background. We performed aperture photometry on the GALEX data using full-sized, half-sized, and quarter-sized versions of the apertures reported by \cite{lee2009_tables}. We found that the half-sized apertures minimized cosmic X-ray background (CXB) contamination, closely matched the effective radius of most galaxies in our sample, and that they are similar to the apertures used in L21. We therefore adopt the half-sized apertures as our galaxy footprints.

NGC 5408 and ESO 495-G021 (He2-10) have no available UV-observations with GALEX. We use the 2MASS apertures provided in L21 since these galaxies are sufficiently bright in the $K_s$-band ($9.00$ and $11.39$ mag, respectively, \cite{Skrutskie2006AJ}). The positions and X-ray photometry for sources within and surrounding the L21 apertures are presented in Table A1 of \cite{L21}. To identify sources that fall within our galaxy aperture, we apply a positional filter on the point sources flagged in the L21 table as 1, 3, and 5 which correspond to point sources within the L21 aperture, sources outside the L21 aperture, and sources that are beyond 1.2 times the galaxy boundaries, respectively. We report our adopted apertures in Table \ref{table:overview} and display the apertures of Chandra+HST galaxies in Figure \ref{fig:galaxies} (magenta ellipses).

\subsection{Star Formation Rates}
\label{subsec:sfr}

To ensure consistency and take into account the new apertures discussed in Section \ref{subsec:aperture}, we recalculate the FUV derived SFRs of each galaxy using GALEX data. To achieve this, we measure the total flux enclosed within our FUV galaxy apertures and correct for Milky-Way foreground extinction using E(B-V) colors from \cite{schlegel1998ApJ} (as reported in \cite{lee2009_tables}). The FUV fluxes are then converted to luminosities using the distance to the galaxy, which in turn is used to measure the SFR. We expect internal FUV extinction in the dwarf galaxies to be small because our combined sample is predominantly comprised of blue dwarf galaxies with negligible dust. We use the scaling relation given by \cite{lee2009_tables} to convert FUV luminosity to SFR:

\begin{equation}
SFR = 1.4 \cdot 10^{-28} \cdot L_\nu(UV)
\end{equation}

Where $L_\nu(UV)$ is the total FUV luminosity density ($\ergs\ {\rm Hz}^{-1}$) enclosed. For the two galaxies with no GALEX data, we used the original SFR values from L21.

\subsection{X-ray Photometry and Source Catalogs}
\label{subsec:x-rayphoto}

The Chandra ACIS data were retrieved from the Chandra archive and homogeneously reprocessed using {\tt chandra\_repro} as part of the {\tt CIAO} package version ${\rm 4.14}$ along with associated calibration files CALDB ${\rm v4.9.6}$. As is standard, we applied the latest bad pixel masks and identified afterglow effects to create new masks, as well as applying the very faint processing flag to further clean the particle background for those observations performed in VFAINT mode to create the final reprocessed Level 2 events files and their associated response files. X-ray photometry was further carried out using \texttt{CIAO}. We started by constructing images in the $0.5-7.0$ keV band from the reprocessed Level 2 events files for the 31 dwarf galaxy candidates in the Chandra+HST sample. For each of the galaxies, we used the \texttt{CIAO} \texttt{fluximage} script to create exposure-corrected images using the reprocessed ACIS graded Level-2 event files in good time intervals (GTI) along with the associated aspect solution files and bad pixel masks. The PSF radius was computed using the \texttt{mkpsfmap} script with the PSF energy set to $1.4967$ keV and the encircled counts fraction (ECF) set to $0.9$. Using the resulting flux image and PSF map as inputs, we ran \texttt{wavdetect} with wavelet scales of $1.0$, $1.414$, $2.0$, $2.828$, $4.0$, $5.657$, $8.0$ pixels and \texttt{sigthresh} set to $10^{-6}$. This step produced our X-ray source photometry (counts) and catalog (positions). 

For each galaxy, we filter out X-ray sources outside of the ACIS chip and the galaxy's aperture. In particular, for ACIS-S observations we filtered out sources that are not in the same chip as the target galaxy, while for front-illuminated ACIS-I observations we kept sources within the four chips (including the gaps). We visually inspected each galaxy to identify the presence of X-ray features that we deemed indicative of an AGN at the center of the flux images. UGC 5139, NGC 5253, NGC 4490, NGC 4605, and NGC 2500 were flagged as having possible AGN in the form of an X-ray point source coincident with their centers. As such, these suspected AGN point sources were filtered out. 

We used the PIMMS (Portable, Interactive Multi-Mission Simulator) to calculate the ACIS counts-to-flux conversion factors for each galaxy based on its observation cycle. The input energy range was set to $0.5-7.0$ keV for all galaxies. Assuming a power-law spectral model we set $\Gamma=1.7$ and estimated the galactic neutral hydrogen column density for each galaxy using HEASARC's software. We then calculated the point-source fluxes by multiplying the counts from \texttt{wavdetect} by the conversion factor we obtained from PIMMS. Furthermore, we calculate the $0.5-7.0$ keV X-ray point-source luminosities ``L'' using the distances to the host galaxy.

\subsection{Completeness Corrections}
\label{subsec:rf}

\begin{figure}[t]
\includegraphics[width=8cm]{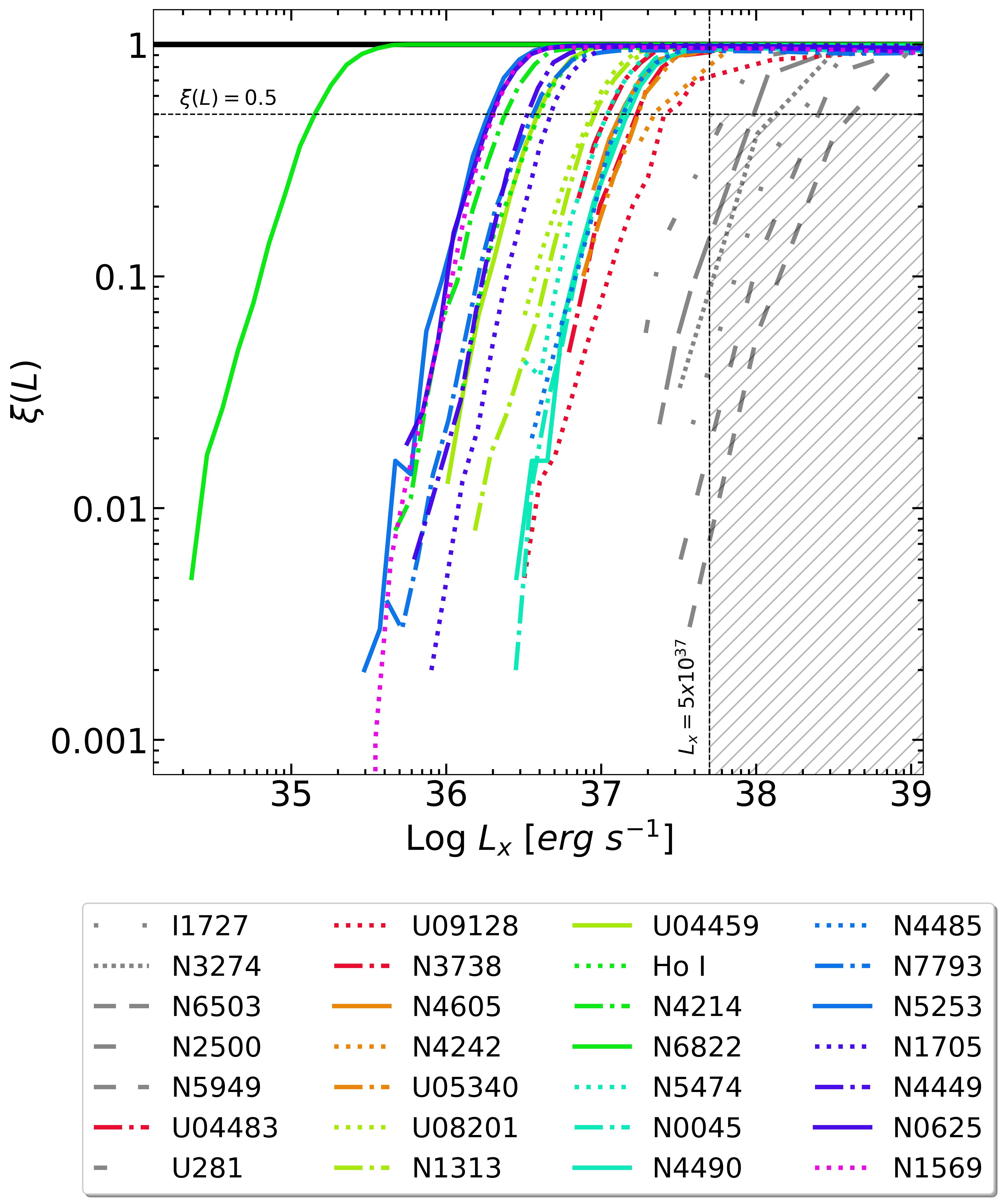}
\centering
\caption{\label{fig:rf} 
Recovery functions within the apertures of the Chandra+HST sample of galaxies. Sub-array mode galaxies are not included in this figure. Galaxies are colored according to relative exposure times (i.e by ranking). The Recovery functions were computed by injecting simulated X-ray sources into Chandra images and measuring the fraction of the synthetic sources recovered during the source detection stage. The minimum \Lx in the simulations is set by the luminosity, at the galaxy's distance, that correspond to 4 counts. The gray vertical and horizontal lines denote a log luminosity of $37.5$ ergs s-1 and $50\%$ recovery respectively. If the $\xi(L)$ profile falls below $50\%$ before $37.5$ ergs s-1, we do not include that galaxy in our analysis (Recovery functions of rejected galaxies are plotted as gray lines). We note that the recovery function of NGC 6822 extends to relatively lower luminosities despite its exposure time because of its proximity to the Milky Way.}
\end{figure}

The recovery function, denoted by $\xi(L)$, is a statistical measure of the fraction of sources that can be detected at a given X-ray luminosity while addressing source crowding and the galactic backdrop's impact on point source detection. As such, we compute the recovery function of each galaxy in our sample by simulating a mock image and running our detection algorithm on the simulated dataset. We start with the original Chandra dataset as a base and use \texttt{CIAO}'s \texttt{simulate\_psf} function to inject $100$ additional synthetic sources into the galaxy aperture within the same point source luminosity bin. \texttt{simulate\_psf} accounts for streaks and PSF distortions. We run \texttt{wavdetect} on the simulated dataset and measure $\xi(L)$ as the fraction of the injected sources that we are able to recover. We repeat this for each luminosity bin $10$ times (a total of $1000$ synthetic sources per luminosity bin per galaxy). We defined $\logL\ (\ergs) = 39$ as our maximum recovery function luminosity, while the lower bound is set to one or two bins below the $4$ count limit. We sample in intervals of $0.5$ dex for luminosities in the range $\logL\ (\ergs) = 37-39$ and $0.1$ dex for $\logL\ (\ergs) < 37$. We use smaller energy bins for $\logL\ (\ergs) < 37$ to sufficiently sample the curvature of the recovery function as it begins to fall from unity to zero with decreasing luminosity. The recovery functions of all the galaxies in our sample are displayed in Figure \ref{fig:rf}.

To ensure the fidelity of our recovery functions, we take the placement of the synthetic sources and the size of each galaxy into consideration. All of the injected point sources are placed such that they are at least one half PSF radius away from each other, even if they are not injected into the same simulated image. This ensures that no two point sources are probing the same area and reduces the chance of source confusion caused by the synthetic sources overlapping with each other. Note that we do allow synthetic sources to overlap with real sources and incorporate such confusion into the recovery function. Since our sample contains dwarf galaxies, their sizes may be too small to allow for a meaningful measurement of $\xi(L)$. To address this, we set the semi-major and semi-minor axes of the galaxy (i.e our sampling area) to a minimum of $3$ arcmin. We also restrict the sampling area to be within the ACIS chip projected footprint (except for gaps in ACIS-I images).

\section{Modeling}
\label{sec:modeling}

In this section, we discuss forward modeling of the X-ray luminosity functions. Following L19 and L21, we model the observed X-ray luminosity distributions by taking into account the low-end completeness of the observations (Section \ref{subsec:rf_model}), contributions from the cosmic X-ray background (CXB, Section \ref{subsec:cxb}), and high and low mass binaries (Section \ref{subsec:xrb_model}). By modeling each component of the luminosity function separately and combining them into a compound model, we can represent and fit the observed data. The CXB for Chandra has been modeled by \cite{kim_cxb_2007} and the completeness of each observation is modeled using simulated recovery functions (see Section \ref{subsec:rf}). We find the contributions of LMXBs, which scale with stellar mass, to be negligible for our sample of galaxies and use the L19 LMXB model to represent their magnitude. The last component is the HMXB component, which scales with SFR. We use the Astropy \citep{astropy22} {\tt models} sub-package to implement individual and compound models. We describe each model component in detail below. 

\subsection{Recovery Function Model}
\label{subsec:rf_model}

We model the fraction of point sources recovered as a function of luminosity \Lx (i.e completeness) using the recovery functions from L21 and the simulated recovery functions in Section \ref{subsec:rf}. We implement the model by using the tabulated recovery fractions as look-up tables, interpolating between data points when necessary. For luminosities that correspond to $4$ counts or less, we enforce a recovery fraction of $0$. For luminosities above the range of the simulated recovery function, we use a recovery fraction of $1$.  

\subsection{Cosmic X-ray Background}
\label{subsec:cxb}

The CXB is the combined X-ray flux from all distant bright X-ray sources, such as AGNs, that may be confused with relatively fewer luminous sources within a nearby galaxy. To model the contribution of such sources, we use the extragalactic X-ray point source number counts from \cite{kim_cxb_2007}. They provide broken power law models, given in Equation \ref{eq:kim1}, for the ChaMP+CDFs \citep{champ, cdfs} number counts per unit area, with parameters listed in their publication's Table 4 for each galaxy. We scale this model by the area of our apertures and convert the CXB flux values to luminosities using the distance to the galaxy. To cover the same energy range as our observations, we use the $0.5--8$~keV ChaMP+CDFs models. Given a power-law photon index of $1.4$, the difference between the $0.5-7$~keV and $0.5-8$~keV energy ranges is negligible ($10\%$). Our adopted model for the flux ($S$) dependant CXB number counts can be expressed as:

\begin{equation} \label{eq:kim1}
    \frac{dN_{{\rm CXB}}}{dS} = \begin{cases} 
      K \left(\frac{S}{S_{\rm ref}}\right)^{-\gamma_1} & S < S_b  \\
      K \left(\frac{S_b}{S_{\rm ref}}\right)^{(\gamma_2-\gamma_1)}\left(\frac{S}{S_{\rm ref}}\right)^{-\gamma_2}& S \ge S_b \\
   \end{cases}
\end{equation}

where $\gamma_1$ and $\gamma_2$ are the broken power-law slopes, $K$ is a normalization constant and $S_{\rm ref}$ is a normalization flux set to $10^{-15}~ \ergs {\rm cm}^{-2}$, and $S_b$ is the break flux at which the slope changes. The differential number count (${dN_{{\rm CXB}}}/{dS}$) as stated in Equation \ref{eq:kim1} is in units of $10^{-15}\ \text{deg}^{-2}$.

\subsection{LMXB and HMXB Models}\label{subsec:xrb_model}

We model the LMXB contributions as a function of stellar mass utilizing the L19 broken power-law model (Equation 12 in L19) with a cutoff luminosity. We use parameter values obtained by the L19 full sample (Table 4 of L19, column 4). We do not fit any parameters for the LMXB at any point because the contributions from LMXBs are negligible. 

For the HMXB, there are two models available. The first model is an SFR-normalized single power-law model from L19, with a cutoff luminosity. The L19 HMXB differential number counts as a function of X-ray luminosity has the following form:

\begin{equation} \label{eq:L19B(L)}
B(L) = K_{\text{HMXB}, 38} \begin{cases} 
       L_{38}^{-\gamma} & L < L_{c} \\
       0 & L \ge L_{c} \\
   \end{cases}
\end{equation}

\begin{equation} \label{eq:L19HMXB}
    \frac{dN_{\text{HMXB}}}{dL} = \text{SFR} \cdot B(L)
\end{equation}

where $\text{SFR}$ is the host galaxy star formation rate, $K_{\text{HMXB}}$ is a normalizing constant, and $\gamma$ is the power-law slope. Following L19, we take $L$ to be in units of $10^{38}\ \ergs$ (denoted by $L_{38} = L/10^{38}$) for the HMXB and LMXB models. The only component that varies across galaxies in this model is the SFR, with $B(L)$ being universal for all galaxies in a given metallicity bin.

The L21 HMXB model introduces a metallicity dependence to the differential number counts (Equations 1, 2, and 3 in L21). The function has the form of a broken power-law with an exponential cutoff. The high-luminosity power-law exponent, as well as the cutoff luminosity, have metallicity dependencies. When we include the L21 model for comparison, we do not refit any parameters but use the values reported in Table 2 of L21.

\subsection{Compound Model}\label{subsec:compound_model}

Following L19 and L21, we combine all the elements described above to forward model the underlying HMXB luminosity function (Equation \ref{eq:L19HMXB}) implied by the ensemble measured luminosity distributions, accounting for incompleteness, CXB, and the LMXB population. The underlying HMXB luminosity function is weighted by the SFR of each galaxy contributing to the ensemble. To account for completeness, the contribution of each galaxy is weighted by its recovery function at each luminosity. This provides us with a model that represents the observed total number counts in each \Lx bin ($\Delta n$):

\begin{equation} \label{eq:deltan}
    \Delta n = \sum_{i}^{N_{gal}} \xi_i(L) \Delta L {\left[\frac{dN_{\text{HMXB}, i}}{dL} + \frac{dN_{\text{LMXB}, i}}{dL} + \frac{dN_{{\rm CXB, i}}}{dL} \right]}
\end{equation}

Since we are interested in modeling the HMXB contributions, we focus on the SFR independent component $B(L)$ (see Equation \ref{eq:L19B(L)}). We rearrange Equation \ref{eq:deltan} to separate $B(L)$ from the rest of $\Delta n$ as follows:

\begin{equation}
    \Delta n = B(L) \cdot C_1(L) + C_2(L)   
\end{equation}

Where:

\begin{equation}
    C_1(L) = \Delta L \sum_{i}^{N_{\text{gal}}} \xi_i(L) \cdot \text{SFR}_i
\end{equation}

\begin{equation}
    C_2(L) = \Delta L \sum_{i}^{N_{\text{gal}}} \xi_i(L) {\left[\frac{dN_{\text{LMXB}, i}}{dL} + \text{CXB}_i(L) \right]}
\end{equation}

Note again that $B(L)$, which is scaled by the SFR of each individual galaxy, is an underlying model across all galaxies. $C_1(L)$ and $C_2(L)$ are components that scale with SFR, projected area, and stellar mass. Since these components do not contain fitting parameters, we can tabulate the values of $C_1(L)$ and $C_2(L)$ as a function of luminosity bins and fit for $B(L)$. 

\section{Luminosity Function Fits}\label{sec:luminosity_fits}

In this Section we provide information on how we fit the compound model, discussed in Section \ref{subsec:compound_model}, to the observed luminosity functions of sub-samples of the final sample (see Section \ref{subsec:final_sample}). We summarize our results and compare them to past models in Section \ref{subsec:fitting_results}. 

\subsection{Metallicity Dependence of Luminosity Function}\label{subsec:bins}

The sub-samples of the final sample are differentiated based on two factors: (1) metallicity bins with \logoh values of $7.7-8.3$ (low), $8.3-8.9$ (high) and $7.7-8.9$ (full), and (2) whether the sub-sample contained galaxies that were targeted for ULXs. Figure \ref{fig:fits} shows the observed X-ray luminosity functions for the resulting five unique sub-samples\footnote{The $8.3-8.9$ metallicity bin does not contain any galaxies that were specifically targeted for ULXs.}.

\subsection{Observed X-ray Luminosity Functions}
\label{subsec:Lx_bins}

The observed X-ray luminosity functions are constructed by aggregating the luminosities of all X-ray point sources within all galaxy apertures in a given metallicity bin. X-ray point source luminosities are binned into intervals similar to L21. The luminosity bins range from $\logL\ (\ergs) = 35-41.7$ and each bin spans $0.1$ dex, resulting in $78$ bins. When inferring models, the log mid-points of the bins are used. The final sample luminosity functions for each metallicity bin are plotted as black points with error bars in Figure \ref{fig:fits}. The one-sigma Poisson errors for the number of sources in each luminosity bin are calculated according to \cite{nasa_poisson}.

\subsection{Model Fitting}
\label{subsec:model_fitting}

As discussed above, the observed luminosity function for each sub-sample is constructed by consolidating all X-ray point sources in all galaxies within a metallicity bin. For each metallicity bin, we limit the range of L that we fit to the highest and lowest L bins that contain at least a single source. We initialize the compound model discussed in Section \ref{subsec:compound_model} using the full sample parameter values in L19 and fix the value of $L_c$ to $10^{40.7}~\ergs$. The values of $C_1(L)$ and $C_2(L)$ are tabulated at the mid point of each luminosity bin. We use a Levenberg-Marquardt algorithm (Astropy's \texttt{LevMarLSQFitter}) to fit $K_{\text{HMXB}}$ and $\gamma$, which are parameters of $B(L)$, in the compound model (see Equations \ref{eq:L19B(L)}, \ref{eq:L19HMXB}, and \ref{eq:deltan}). The quality of our models' representation of the data was evaluated using the C-statistic, as described in Section \ref{subsec:goodness}.

We account for the uncertainties in SFR by implementing a Monte Carlo sampling technique, which consists of $10^4$ iterations. In each iteration, the luminosity function is refit with the SFR for each galaxy drawn from a normal distribution of SFR with a mean equal to the measured SFR and a standard deviation equal to the corresponding one-sigma error. If the galaxy does not have FUV SFR measurements, we use half the SFR value as the standard deviation to reflect the relatively higher degree of uncertainty. We limit the range of allowed SFRs to 5 standard deviations from the mean and enforce a minimum SFR of $0.0001$. The mean of the parameters resulting from all the iterations is taken as our final result. It is worth mentioning that the fluctuations in the SFR do not result in substantial propagated uncertainties in the fitted parameters. The parameter errors of each fit (at each iteration) are estimated by taking the diagonal of the covariance matrix from the Levenberg-Marquardt optimizer for the best SFR.

Consistent with L19, we adopt a cutoff luminosity ($L_c$) from their comprehensive sample fit. However, it's worth noting that this cutoff luminosity, while empirically motivated, may have significant uncertainty. This choice of fixed $L_c$ could bias our fit. Thus, to validate its impact, we conduct a test to estimate the number of sources that our best fit predicts above this cutoff. We integrate the fitted luminosity functions above $L_c$ and find that two or fewer sources are expected. This negligible result alleviates concerns about our fit converging on an unphysically flat slope due to artificial truncation.

\subsection{Goodness of Fit}
\label{subsec:goodness}

Following L19, we evaluated the goodness of fit for each of the metallicity bins on a global basis using a modified C-statistic \citep{Cash79, Kaastra2017}:

\begin{equation}\label{eq:ctat}
    C = 2 \sum_{i=1}^{n_b} M_i - N_i + N_i \ln{\left(N_i/M_i\right)}
\end{equation}

where $C$ denotes the C-statistic corresponding to a particular metallicity bin. The sum of the statistic is calculated by iterating through the luminosity bins. $n_b$ is the total number of luminosity bins, whereas $M_i$ and $N_i$ represent the model value and the observed counts for the $i^{th}$ bin, respectively. We mask out bins where the model is equal to zero ($C_i = 0$ if $M_i = 0$). For bins in which the observed number of sources is zero, we set the logarithmic component to zero for that term ($N_i \ln{\left(N_i/M_i\right)}= 0$ if $N_i = 0$).

We adopted the methods outlined by \cite{Kaastra2017} to compute the expected C-statistic ($C_{exp}$), along with its variance ($C_{var}$), based on Poisson statistics. Subsequently, we evaluated the null-hypothesis probability ($P_{null}$) as follows:

\begin{equation}
    P_{null} = 1 - \rm{erf} \left(\sqrt{\frac{(C - C_{exp})^{2}}{C_{var}}} \right)
\end{equation}

Where $P_{null}$ is the null-hypothesis probability, C is the C-statistic measured from the data following Equation \ref{eq:ctat}, and $\rm{erf}$ is the error function. Models with $P_{null} < 0.001$ can be statistically rejected with greater than $99.9\%$ confidence. Utilizing a modified C-statistic, we assess the goodness of fit across various metallicity bins and this threshold ensures the reliability of the models we test.

\subsection{Fitting Results and Comparisons}
\label{subsec:fitting_results}

\begin{figure*}[t]
\includegraphics[width=\textwidth]{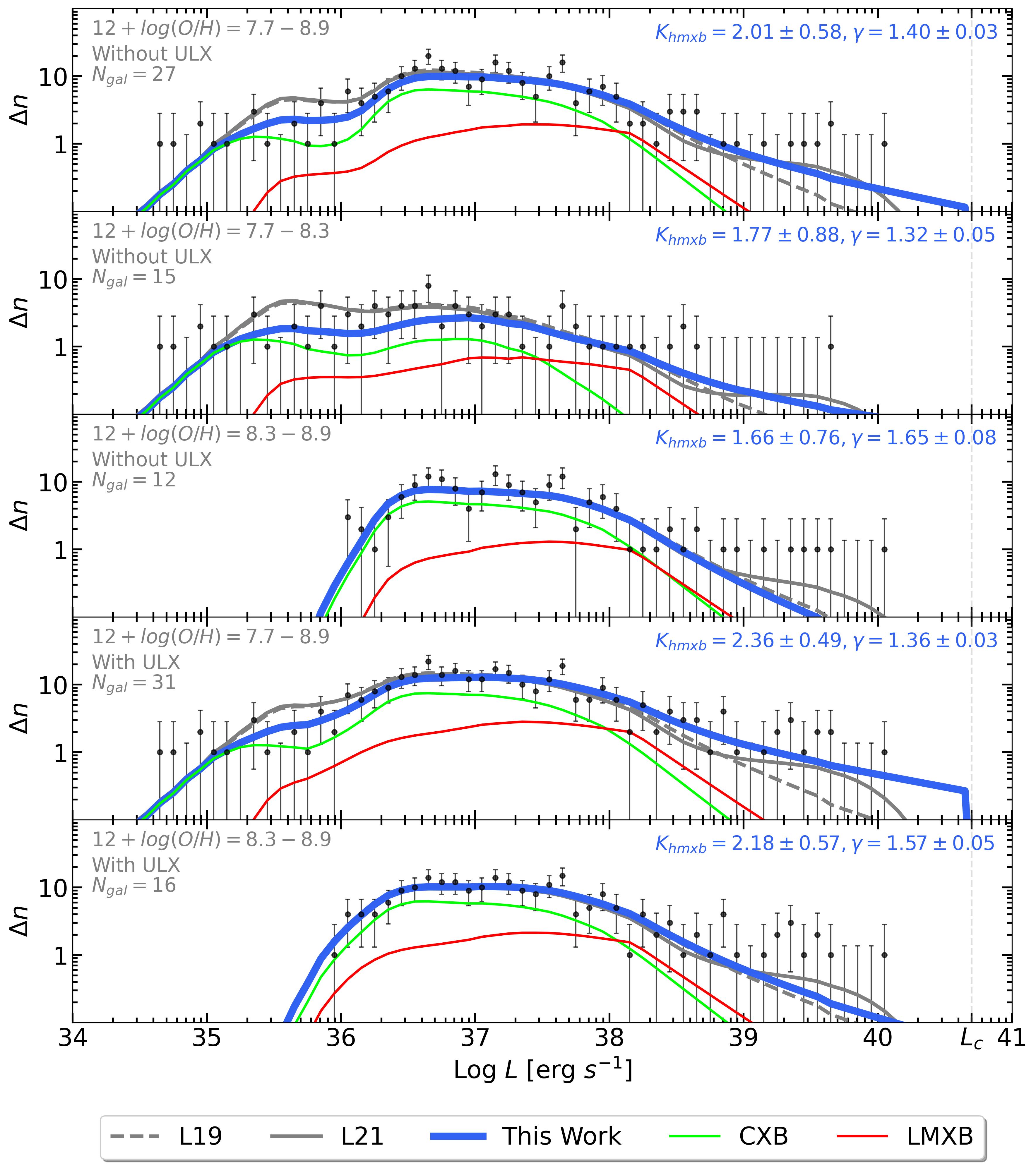}
\centering
\caption{\label{fig:fits} 
This figure shows models that were fit to data in three gas-phase metallicity bins. All panels show the mass and distance filtered final sample of galaxies. The first three panels show samples without galaxies that were targeted for ULXs while the last two include them. The metallicity range and the number of galaxies in each panel is given the upper left corner. The first and fourth panels represent the entire dwarf sample within the metallicity range of this study (i.e $\logoh = 7.7-8.9$). Observed distributions of X-ray point source luminosities ($\Delta n =\ $dN/dL) for galaxy sub-samples are plotted with one sigma Poisson error bars (black). Recovery functions have been applied to all models to match the expected completeness of the data. The red and green lines model LMXB and CXB contributions respectively. The gray dashed, solid lines, and blue lines show the outputs of the combined (HMXB + LMXB + CXB) models for L19, L21, our study. The numeric values of the fitted amplitude ($K_{hmxb}$) and slope ($\gamma$) are given at the top right corner of each panel. }
\end{figure*}

\begin{figure}[t]
\includegraphics[width=8cm]{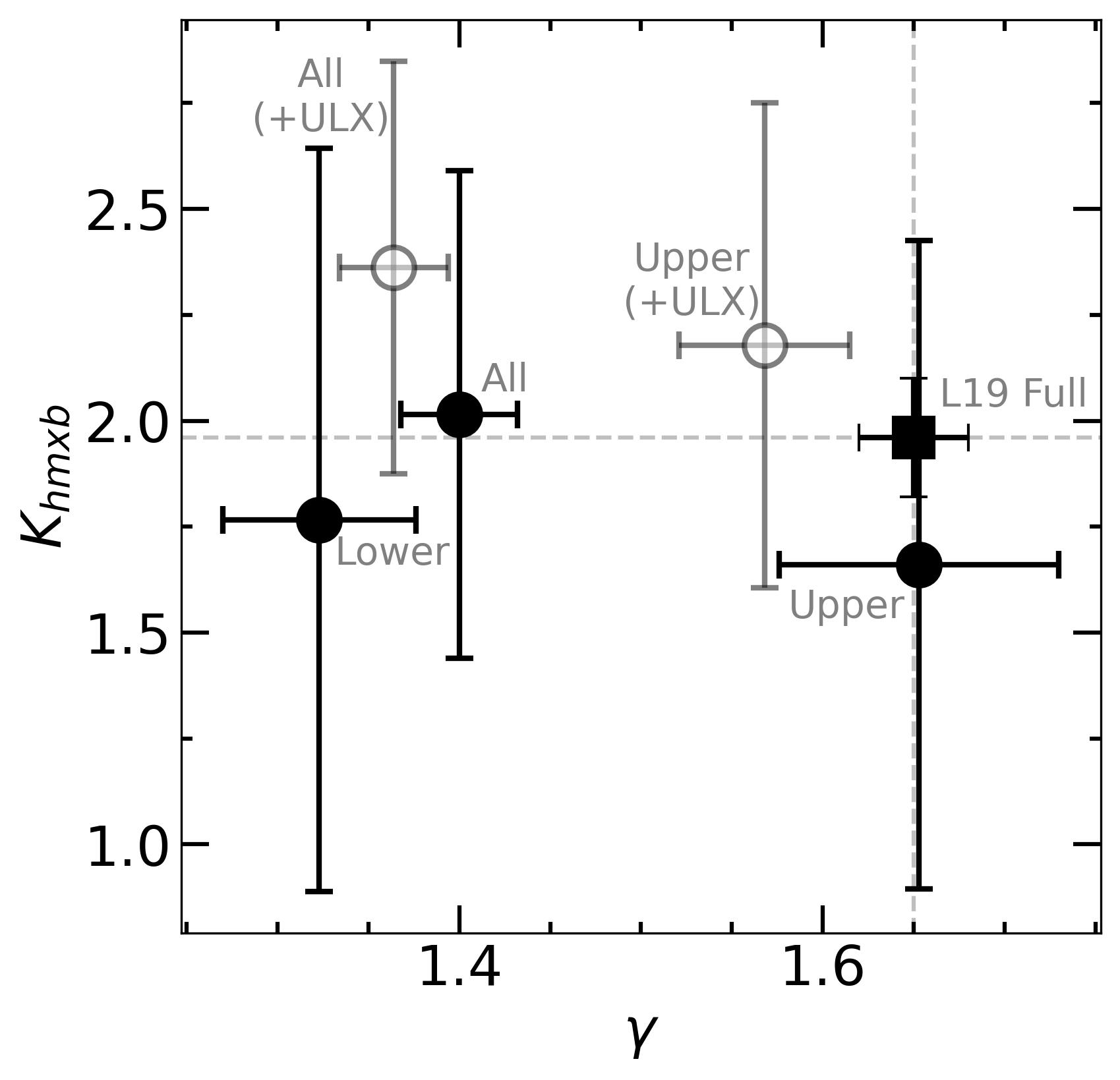}
\centering
\caption{\label{fig:fit_values} 
Parameter fits and associated one sigma errors for five sub-samples compared to L19 parameters. The square marker denotes the L19 fitted parameters and one-sigma errors for their full sample. The gray dashed lines mark the L19 parameter values to help with comparison. Our sub-samples without ULX galaxies are marked with filled circles while sub-samples with ULX galaxies are marked with empty circles. The metallicity bins of the sub-samples are labeled by name, where All=$7.7-8.9$, Lower=$7.7-8.3$, Upper=$8.3-8.9$.}
\end{figure}

\begin{flushleft}
\begin{table*}[htbp]
\footnotesize
\centering
\begin{tabular}{l*{8}{c}}
\hline 
\multicolumn{9}{c}{Model Parameters and Goodness of Fit Evaluation} \\ 
\hline 
\multicolumn{1}{c}{12 + log[O/H] Bin} & \multicolumn{1}{c}{$K_{HMXB}$} & \multicolumn{1}{c}{$\gamma$} & \multicolumn{1}{c}{C} & \multicolumn{1}{c}{$C_{exp}$} & \multicolumn{1}{c}{$C_{var}$} & \multicolumn{1}{c}{$P_{null}$} & \multicolumn{1}{c}{$P_{null,L19}$} & \multicolumn{1}{c}{$P_{null, L21}$}\\ 
\multicolumn{1}{c}{(1)} & \multicolumn{1}{c}{(2)} & \multicolumn{1}{c}{(3)} & \multicolumn{1}{c}{(4)} & \multicolumn{1}{c}{(5)} & \multicolumn{1}{c}{(6)} & \multicolumn{1}{c}{(7)} & \multicolumn{1}{c}{(8)} & \multicolumn{1}{c}{(9)}\\ 
\hline
7.7-8.9 & $2.02\pm0.57$ & $1.40\pm0.03$ & 71 & 61 & 100 & 0.316 & 0.013 & 0.05\\ 
7.7-8.3 & $1.77\pm0.88$ & $1.32\pm0.05$ & 54 & 57 & 90 & 0.786 & 0.383 & 0.435\\ 
8.3-8.9 & $1.66\pm0.77$ & $1.65\pm0.08$ & 53 & 40 & 74 & 0.107 & 0.324 & 0.739\\ 
7.7-8.9 (+ULX) & $2.36\pm0.49$ & $1.36\pm0.03$ & 64 & 64 & 105 & 0.982 & $>$0.001 & 0.023\\ 
8.3-8.9 (+ULX) & $2.18\pm0.57$ & $1.57\pm0.05$ & 53 & 47 & 83 & 0.53 & 0.138 & 0.739\\ 
\hline 
\end{tabular}
\caption{Fitting results and goodness of fit tests. (1) Gas-phase metallicity. (2-3) The values of the model parameters obtained after fitting. (4) Cash statistics value for fitted parameters. (5-6) Expected value and variance of the Cash statistics for the fitted model. (7) Null hypothesis probability for the fitted model. (8-9) Null hypothesis probability for L19 and L21 models.}\label{table:fitting}
\end{table*}
\end{flushleft}

The results of our fits are summarized in Table \ref{table:fitting}, and the corresponding figures are presented in Figures \ref{fig:fits} and \ref{fig:fit_values}. The differential number counts of X-ray point sources as a function of luminosity are provided in Figure \ref{fig:fits}, grouped as described previously, and the best-fit parameters for the luminosity function models are provided in \ref{fig:fit_values}.

Our analysis of the full metallicity bin reveals a power-law slope of $\gamma = 1.40 \pm 0.03$. When fitting the same metallicity bin, including galaxies targeted for ULXs, we find a slightly shallower power-law slope of $\gamma = 1.36 \pm 0.09$. However, this difference is not statistically significant, indicating that the presence of ULX targeted galaxies does not have a meaningful effect on the overall slope of the dwarf HMXB population.

We examine the metallicity dependence of the HMXB luminosity function by fitting and comparing sub-samples of galaxies binned by metallicity ranges (Section \ref{subsec:bins}). We find that the low metallicity bin is best fit by a shallower slope than the full sample, while the high metallicity bin results in a steeper slope. The shallower slope of the low metallicity bin implies a relative excess of high luminosity sources in that bin. Interestingly, only the higher metallicity bin includes ULX-targeted galaxies. Similar to the full metallicity sample, ULX-targeted galaxies in the high metallicity sample have slightly shallower luminosity functions slopes, but the difference is not statistically significant. An unbiased sample is likely to provide results that lie somewhere between the ULX-included and ULX-excluded sample results.

In Figure~\ref{fig:fit_values}, we compare the best-fit parameters to the L19 results, which utilizes a broader range of stellar masses, inclusive of galaxies larger than dwarfs. We find that dwarf galaxies exhibit a shallower slope of $\gamma \sim 1.40 $ compared with the full L19 sample, $1.65 \pm 0.03$, equating to an $\sim 8.0 \sigma$ deviation, suggestive of a mass and/or metallicity dependence on the power-law slope. By contrast, we find no evidence for a dependence on the normalization parameter $K_{\rm HMXB}$. We also note that the systematic offsets in the normalization parameter due differences in SFR measurements are are found to be minimal. Specifically, the average ratio of our measurements to those in L21 is $1.07$, with a standard deviation of $0.5$, with a maximum deviation of $\sim 2.4$. We test how well these models describe the HMXB population by taking into consideration the null-hypothesis probability derived from their C-stats. We find that all models result in acceptable fits, with the exception of the L19 model in the case of the full metallicity bin with ULX galaxies.

\subsection{Observed \LxOverSFR}
\label{subsec:observed_LxOverSFR}

In addition to looking at the luminosity function for X-ray binaries across all the galaxies in the sample, we can also explore how the X-ray luminosity from individual sources varies with the gas-phase metallicity. Since we know that SFR is the primary driver of the $L_X$ from high-mass X-ray binaries, we must normalize each $L_X$ by the SFR in that galaxy in order to isolate any additional correlation with metallicity.  

\begin{figure}[t]
\includegraphics[width=8cm]{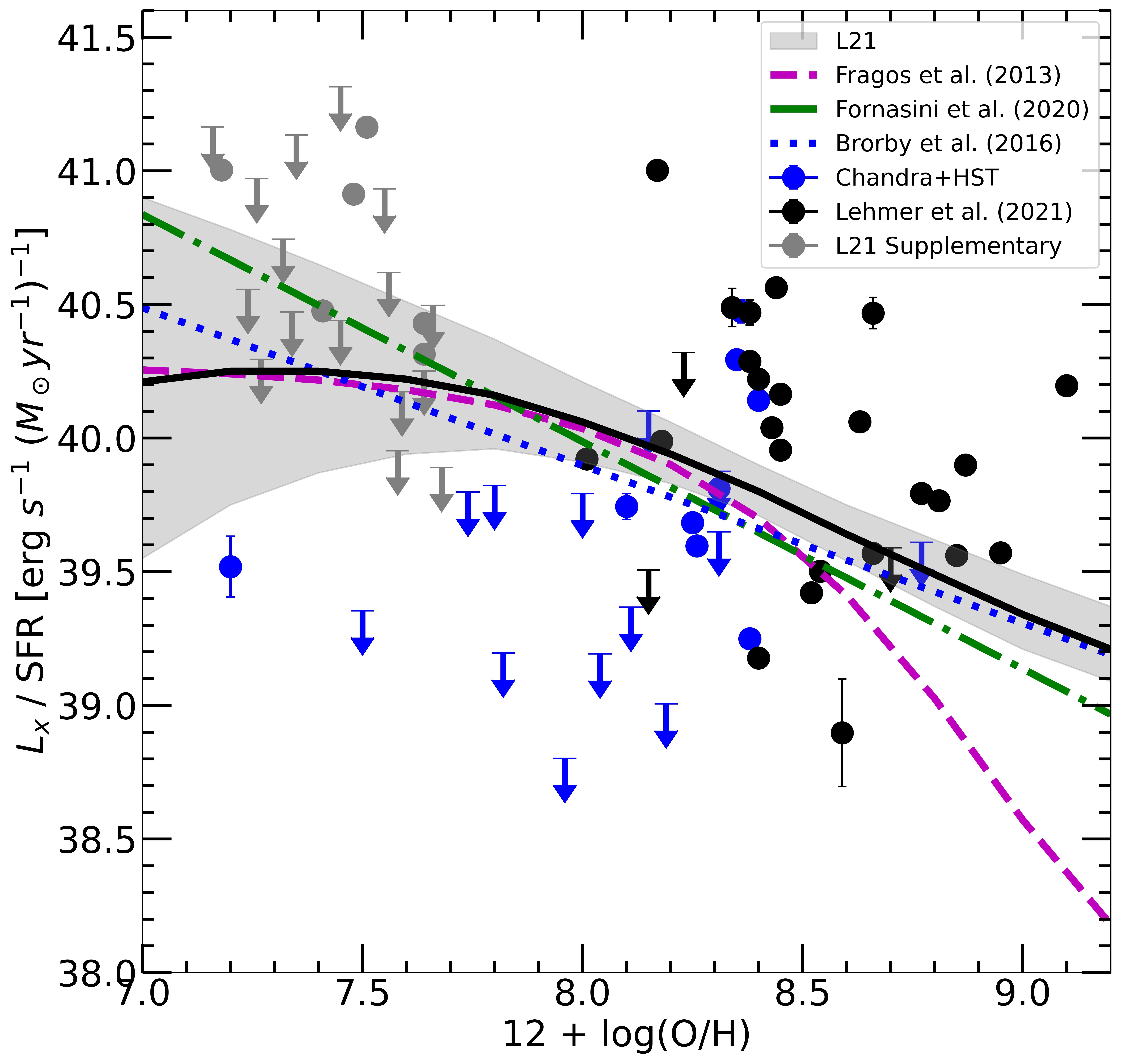} 
\centering
\caption{\label{fig:lx_sfr_log_oh}
\LxOverSFR (HMXB) vs metallicity relation for galaxies in the Chandra+HST and all L21 samples. Dwarfs from the Chandra+HST sample are plotted in blue and L21 galaxies are plotted in black. Galaxies from the L21 supplemental sample, which are compact dwarfs, are plotted in gray. The total HMXB Lx values are calculated by summing the observed Lx and subtracting out the LMXB and CXB contributions after correcting for completeness. Upper limits are used for galaxies with few or no sources using the expected LMXB and CXB contributions.}
\end{figure}

We consider \LxOverSFR as a function of metallicity for individual galaxies in Figure \ref{fig:lx_sfr_log_oh}. For each galaxy in the Chandra+HST sample, the total HMXB X-ray luminosity is computed by summing the observed point source luminosities and subtracting the expected LMXB and CXB contributions. The galaxies with \Lx below the expected LMXB and CXB contributions have upper-limits that are $3 \sigma$ above the expected background from these two sources in that galaxy. 

With our expanded sample of dwarf galaxies, the vast majority of low-mass systems fall well below the predicted \LxOverSFR relations. These low \LxOverSFR values cannot be explained by incompleteness or aperture effects. Many of our deepest exposures have no HMXBs, and even when we double our SFR apertures, we recover the same result. As we will discuss in the next section, this wide range in \LxOverSFR\ is actually expected, because at the low SFR of the dwarfs in the Chandra+HST sample ($0.01-0.1~M_\odot/{\rm yr}$), galaxies cannot fully populate the HMXB X-ray luminosity function, and instead only a very small number of galaxies are expected to harbor a luminous X-ray binary at any given time due to stochastic sampling \citep{Gilfanov2004MNRAS, L21}.

\section{The $L_{\rm X}$-SFR-Metallicity Relation}
\label{sec:lx-sfr-z}

We have introduced an expanded sample of dwarf galaxies with measured X-ray binary luminosity functions, covering a broader range of sSFR and metallicity than prior work. This additional sample shows evidence for an excess of luminous X-ray sources at low metallicity in the stacked luminosity functions, and also highlights the challenges of systematic studies in dwarf galaxies where the typical star formation rates are low. In order to put our results into context, we first review the theoretical reasons to expect a dependence on metallicity. We then contextualize our measured distribution of \LxOverSFR\ with prior work.

\subsection{Model Predictions}
\label{subsec:model_prediction}

Population synthesis models \citep{Dray2006MNRAS, Fragos2013ApJA, Fragos2013ApJB, Madau&Fragos2017ApJ} suggest that the observed scatter in the \Lx-SFR relation can be explained by a secondary dependence of \Lx on metallicity. 

The metallicity dependence is thought to arise because stars with higher metallicity are known to undergo greater angular momentum loss due to stronger winds, which results in orbital expansion that reduces the chance of forming Roche lobe overflow \citep{Fragos2013ApJA}. Moreover, stronger radiatively driven winds in high-metallicity stars cause them to undergo enhanced mass loss before their eventual supernova explosions \citep{Fornasini2020MNRAS}. This process tends to yield compact objects that are relatively lower in mass and less numerous in high-metallicity galaxies, resulting in diminished X-ray luminosities from HMXBs. This anti-correlation causes us to expect lower-metallicity galaxies to have sources that extend to higher \Lx. This trend has been seen in observations \citep[][L19, L21]{Douna2015A&A, Brorby2016MNRAS}, however, the same low-metallicity galaxy sample is in common across nearly all of these studies. These low-metallicity galaxies also have high specific star-formation rates, which correlates with low-metallicity \citep{Mannucci2010MNRAS}. Thus, our Chandra+HST sample both fills in intermediate metallicities and alleviates the strong bias to the highest star-formation rates.

\subsection{SFR Driven Stochasticity}
\label{subsec:stochasticity}

\begin{figure*}[t]
\includegraphics[width=\textwidth]{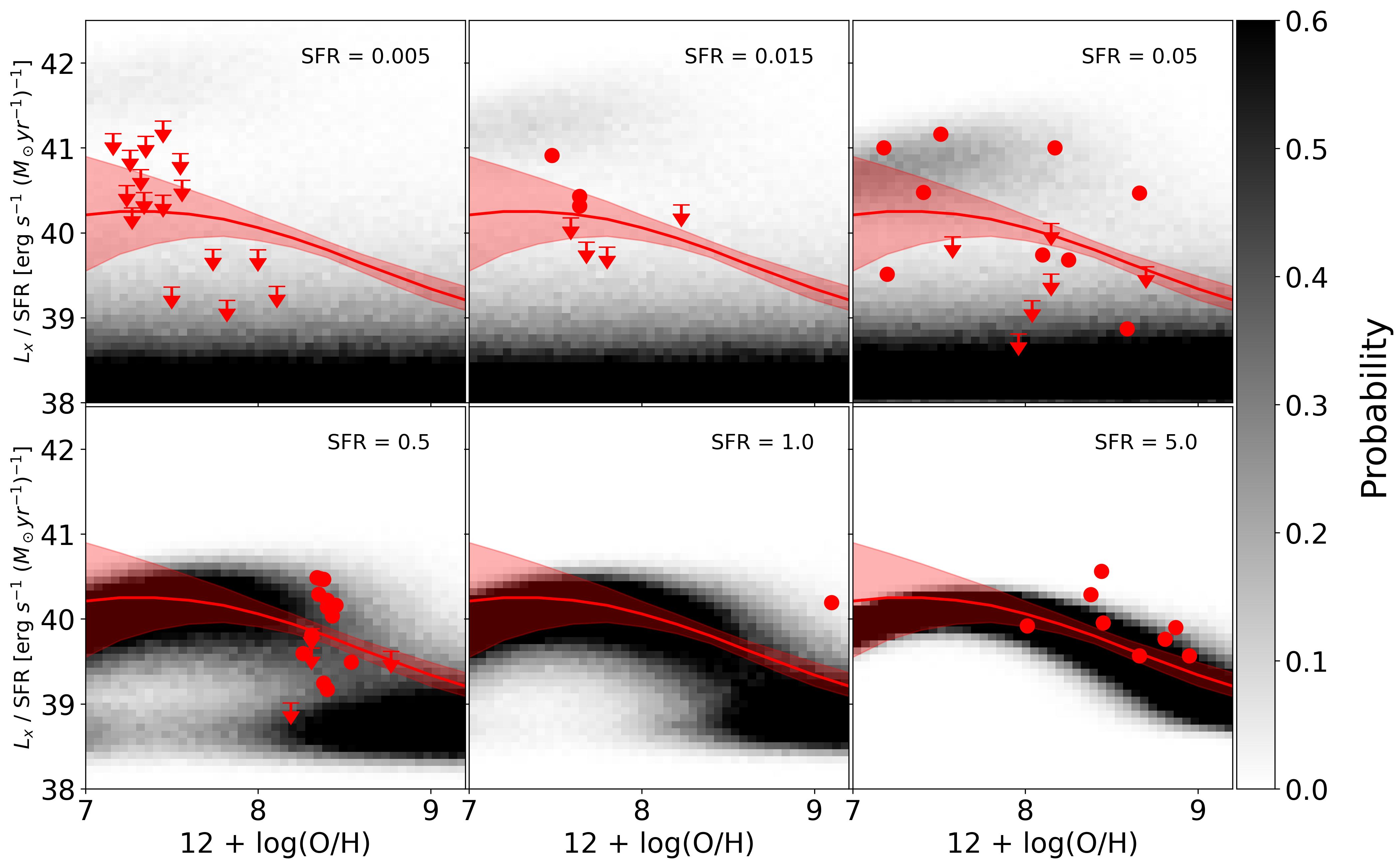}
\centering
\caption{\label{fig:prob_dist} 
We show the probabilistic distribution of \LxOverSFR against \logoh for SFRs and gas-phase metallicities similar to the dwarf samples discussed in this work. The Monte Carlo probabilities were computed, for each SFR and \logoh pair, by using the L21 HMXB model as the probability density function. The six panels correspond to SFRs of $0.005$, $0.015$, $0.05$, $0.5$, $1.0$, and $5.0$ \sfrUnits respectively. The greyscale shows the probability of finding a galaxy with a \LxOverSFR for a given \logoh. The red markers show the observed galaxy \LxOverSFR values and upper-limits, including all galaxies in this work and L21 for reference (not just dwarfs). The red line represents the L21 model prediction for reference. At lower SFRs, the probability of encountering a high \LxOverSFR galaxy is reduced due to stochastic Poissonian sampling, despite a surplus of high \Lx sources in the HMXB luminosity function of galaxies with low \logoh. With increasing SFR, the probability distribution approaches the L21 prediction (red line). Some observed points may hover slightly above or below the probability evaluated at the SFR displayed, this is due to the scatter introduced by the range of galaxy SFRs.
}
\end{figure*}

In agreement with L21, the shallower slopes observed in the stacked luminosity functions (as discussed in Section \ref{sec:luminosity_fits}) imply that flatter luminosity functions are expected (i.e., more sources at higher luminosities) with decreasing metallicity.  At the same time, we observe a considerable spread in \LxOverSFR values for individual sources. 

We can reconcile the shallower slopes in HMXB luminosity function stacks with the large spread in \LxOverSFR by considering the low SFRs of the typical dwarfs in our sample. Specifically, the disparity can be largely attributed to stochastic Poissonian sampling effects that come into play due to the inherently low SFRs in dwarf galaxies. This is due to the fact that \citep[][L21]{Grimm2003MNRAS, Gilfanov2004MNRASb}:

\begin{multline}
\Lx = \int^{L_{\rm up}}_{L_{\rm lo}} L\ B(L)\ dL = K_{\rm HMXB} \int^{L_{\rm up}}_{L_{\rm lo}} L^{-\gamma+1} dL \\
= \frac{K_{\rm HMXB}}{2-\gamma} \left[L_{\rm up}^{2-\gamma} -  L_{\rm lo}^{2-\gamma}\right]
\end{multline}

Where $L_{\rm up}$ and $L_{\rm lo}$ are the upper and lower bounds of the integrated $\Lx$. If the power-law $\gamma$ is less than $2$, then the first term dominates, resulting in a highly stochastic \LxOverSFR distribution through a dependence on the highest X-ray source in the galaxy. In contrast, a $\gamma$ value greater than $2$ results in a stable value. 

\cite{Grimm2003MNRAS} and L21 quantitatively explore the impact of stochasticity at low SFR. In Fig.\ 5 of L21, they present a Monte Carlo simulation of the expected distribution in \LxOverSFR given their fitted HMXB luminosity function. These simulations show an inherent stochastic scatter in \LxOverSFR at low SFR and that the distributions become Gaussian at ${\rm SFR} \geq 2-5\ \sfrUnits$, which are much higher than the typical SFR of dwarf galaxies (see Fig. \ref{fig:sfr_vs_mass}).

To quantify the spread in \LxOverSFR for the Chandra+HST sample, we conducted Monte Carlo simulations utilizing the L21 HMXB luminosity function, treated as a metallicity-dependent probability density function. For each SFR and \logoh pairing, we initially integrated the HMXB number densities (for $\logLx > 35$) to estimate the expected source count. Subsequently, we performed random sampling of the predicted number of HMXBs from the L21 model. This Monte Carlo approach was iterated $10,000$ times, and the mean \LxOverSFR for each SFR and \logoh pair was computed to characterize the distribution. We show the resulting probability distributions in Figure \ref{fig:prob_dist}. The six panels correspond to SFRs of $0.005$, $0.015$, $0.05$, $0.5$, $1.0$, and $5.0$ \sfrUnits respectively. We over-plot the galaxies in Figure \ref{fig:lx_sfr_log_oh} binned by SFR for reference.  

In Figure \ref{fig:prob_dist}, for the lowest SFR bin (${\rm SFR} = 0.005$), we observe a reduced probability of detecting a galaxy with a high \LxOverSFR, a consequence of stochastic Poissonian sampling, despite the over-representation of high \Lx sources in the HMXB luminosity function at low metallicity \logoh. This is because in the case of ${\rm SFR} = 0.005$, we expect small source number counts ($N=2$ sources) which results in near-zero high-luminosity sources populating and dominating the total \Lx of the galaxy, thereby inducing substantial stochastic variability in \LxOverSFR. As the SFR increases (${\rm SFR} = 0.015$ and ${\rm SFR} = 0.05$) the probability of encountering a high \LxOverSFR galaxy is enhanced because of the slightly larger number of sources being sampled from the luminosity functions. If we take a slice at a low \logoh value (vertically) for these SFRs, the \LxOverSFR probability would be a biomodal distribution due to a single source dominating the \Lx of the galaxy. At high SFRs (${\rm SFR} > 1.0$), the larger source number-counts result in the distributions becoming Gaussian, which in turn results in the probabilities approaching the L21 predictions. Thus, at high SFR, the \LxOverSFR approaches the theoretical value with small scatter. The \cite{prestwich_2013} dwarf sample (L21 supplementary) exhibit higher \LxOverSFR because they have much higher specific SFR in general. In particular, six  galaxies (30\% of the supplementary sample) have log \LxOverSFR above $40$ and SFRs that are greater than $0.01$. 

Returning to one of our main motivations for this work, to understand how and when we may use \LxOverSFR\ to search for accreting massive black holes, we have a hopeful message. While it will be necessary to fold in the detailed SFR distributions and (ideally) metallicity distributions for a sample to derive the expected number of detections from HMXBs alone, these distributions mean that the contribution from HMXBs should be low ($\sim 3-4\%$) from dwarfs that lie on the star-forming main sequence and the local mass-metallicity relation.

\section{Summary}

In this paper we present the HMXB X-ray luminosity functions of dwarf galaxies within the local volume, and specifically investigate the role of metallicity in setting the high-mass X-ray binary luminosity function. 

\begin{enumerate}
    \item We introduce the Chandra+HST sample, a selection of local volume dwarf galaxies drawn from the LEGUS and STARBIRDS data-sets. These galaxies have multi-band coverage provided by HST, GALEX, IRAC, and \emph{Chandra}. Combined with 11 local dwarfs from \cite{L21}, we present a total sample of 35 dwarf galaxies spanning a stellar mass range of ($\text{log}\ M_*/M_\odot$) $6.8-9.52$, gas-phase metallicity (\logoh) range of $7.74-8.77$, and a distance range of $0.5-12$ Mpc. The Chandra+HST sample presented here effectively complements the L21 sample by bridging the metallicity gap between their main and supplementary sub-samples, while sampling a wider range of sSFR.
    
    \item We model the HMXB X-ray luminosity functions of local dwarf galaxies across a luminosity range of $\logL = 34.1-41.7$ and a metallicity range of $\logoh = 7.7-8.9$. We find that the power-law slope of the luminosity function is shallower (above 3-sigma) for our local sample compared to the L19 Full Sample, which consists of more massive galaxies with higher metallicities.
    
    \item After dividing our final sample into two metallicity bins, we observe that the lower metallicity bin exhibits a shallower power-law slope, which suggests a relative excess of high \Lx sources as a function of metallicity. This result holds regardless of whether we include or exclude galaxies that were originally targeted for harboring ULXs. 
    
    \item We observe a large spread in \LxOverSFR for galaxies in the Chandra+HST sample. We attribute this deviation to the Poissonian sampling of the luminosity functions that is caused by the relatively lower sSFRs of dwarf galaxies. 
\end{enumerate}

In addition to our results, we expect that star formation histories (SFHs) and galaxy concentration significantly influence the HMXB luminosity distribution and sampling of HMXBs in dwarf galaxies. In particular, galaxies with recent SFHs that deviate from the average dwarf population are prime candidates for further study. Galaxy concentration could potentially introduce variations in the luminosity function of HMXBs due to differences in binary system formation rates, enhanced stellar interactions, and the influence of stellar evolution within densely concentrated regions. Lastly, a larger sample of dwarf galaxies would help in constraining the high log $\LxOverSFR > 40$ regime for the sSFR range explored in this study.

\section{Acknowledgments}
RG and ADG acknowledge funding provided by the National Aeronautics and Space Administration through Chandra Award Number AR2-23011X issued by the Chandra X-ray Center, which is operated by the Smithsonian Astrophysical Observatory for and on behalf of the National Aeronautics Space Administration under contract NAS8-03060. JEG acknowledges support from National Science Foundation awards AAG/1007052 and AAG/1007094. BDL acknowledges support from Astrophysics Data Analysis Program (ADAP) 80NSSC20K0444. This research has made use of new data obtained by the \textit{Chandra X-ray Observatory} and data obtained from the {\it Chandra} Data Archive, as well as software provided by the {\it Chandra} X-ray Center (CXC).

 \software{
\texttt{CIAO} \citep{ciao}
Astropy \citep{astropy22}, 
SciPy \citep{2020SciPy-NMeth}, 
Numpy \citep{2020NumPy-Array},
MatPlotLib \citep{Matplotlib},
}

\bibliography{hmxb_lum_func}{}
\bibliographystyle{aasjournal}

\clearpage

\appendix

In Table \ref{table:appendix} we present a table of point sources within the apertures of the Chandra+HST sample. The full table contains the following columns: Col.(1) Name of the galaxy that the source belongs to. Col.(2-3) Right ascension and declination of the point source. Col.(4) $0.5-7$ keV net counts and 1 $\sigma$ errors. Col.(5) $0.5-7$ flux of the source. 

\begin{flushleft}
\begin{table*}[htbp]
\footnotesize
\centering
\begin{tabular}{l*{5}{c}}
\hline \hline
\multicolumn{1}{c}{Name} & \multicolumn{1}{c}{$\alpha_{J2000}$} & \multicolumn{1}{c}{$\delta_{J2000}$} & \multicolumn{1}{c}{$N_{0.5-7 	\text{keV}}$} & \multicolumn{1}{c}{$Log F_{0.5-7 	\text{keV}}$}\\ 
\multicolumn{1}{c}{(1)} & \multicolumn{1}{c}{(2)} & \multicolumn{1}{c}{(3)} & \multicolumn{1}{c}{(4)} & \multicolumn{1}{c}{(5)}\\ 
\hline
  & H M S & D M S & counts & log erg cm$^{-2}$ $s^{-1}$\\ 
\hline
NGC 0045 & 00 14 03.0 & -23 12 19.1 & $80.41 \pm 12.43$ & $-14.041$\\ 
 & 00 14 03.6 & -23 10 07.1 & $53.38 \pm 8.97$ & $-14.219$\\ 
 & 00 14 06.1 & -23 10 05.8 & $58.29 \pm 7.31$ & $-14.181$\\ 
 & 00 14 11.5 & -23 11 38.5 & $3.26 \pm 7.63$ & $-15.434$\\ 
 & 00 14 04.0 & -23 10 55.5 & $13.28 \pm 1.80$ & $-14.823$\\ 
 & 00 13 58.3 & -23 11 07.3 & $16.58 \pm 3.64$ & $-14.727$\\ 
 & 00 14 00.9 & -23 10 17.0 & $6.78 \pm 3.30$ & $-15.116$\\ 
 & 00 14 01.2 & -23 08 28.0 & $14.83 \pm 5.57$ & $-14.776$\\ 
NGC 0625 & 01 35 03.5 & -41 26 14.2 & $296.07 \pm 17.21$ & $-13.718$\\ 
 & 01 35 07.1 & -41 26 05.4 & $607.84 \pm 24.65$ & $-13.405$\\ 
 & 01 35 07.3 & -41 26 11.2 & $16.79 \pm 3.83$ & $-14.964$\\ 
NGC 1313 & 03 18 07.2 & -66 30 46.0 & $4.27 \pm 9.02$ & $-15.243$\\ 
 & 03 18 06.4 & -66 30 38.6 & $55.67 \pm 4.86$ & $-14.128$\\ 
 & 03 18 05.5 & -66 30 14.9 & $149.06 \pm 17.19$ & $-13.701$\\ 
 & 03 18 18.2 & -66 30 04.2 & $365.14 \pm 2.07$ & $-13.312$\\ 
 & 03 18 18.9 & -66 30 01.5 & $286.30 \pm 7.46$ & $-13.417$\\ 
 & 03 18 20.0 & -66 29 11.1 & $5241.10 \pm 12.21$ & $-12.155$\\ 
 & 03 18 29.5 & -66 28 41.2 & $9.88 \pm 19.11$ & $-14.879$\\ 
 & 03 18 23.7 & -66 28 34.6 & $16.56 \pm 16.92$ & $-14.655$\\ 
 & 03 18 21.2 & -66 28 58.6 & $13.40 \pm 72.40$ & $-14.747$\\  
 \hline
\end{tabular}
\caption{An abbreviated version of the source catalog for the Chandra+HST sample. The full catalog contains 184 sources, and a discritpion is provided in appendix.}\label{table:appendix}
\end{table*}
\end{flushleft}

\clearpage

\end{document}